\documentclass[aps,prb,twocolumn,superscriptaddress,longbibliography]{revtex4-2}

\usepackage{graphicx}    
\usepackage{amsmath}     
\usepackage{amssymb}     
\usepackage{bm}          
\usepackage{placeins}
\usepackage[T1]{fontenc}
\usepackage[latin9]{luainputenc}
\usepackage{babel}
\usepackage{color}

\usepackage{xcolor}
\usepackage[colorlinks=true, citecolor={blue!80!black}, urlcolor={blue!50!black}, linkcolor = {blue!80!black}]{hyperref}

\usepackage{soul}

\begin{document}

\title{Probing spinon interactions in the spin-1 bilinear-biquadratic chain}

\author{Yonatan Lin}
\affiliation{Department of Physics, Technion, Haifa, 3200003, Israel}
\author {Oleg A. Starykh}
\affiliation{Department of Physics and Astronomy, University of Utah, Salt Lake City, Utah 84112, USA}
\author{Anna Keselman}
\affiliation{Department of Physics, Technion, Haifa, 3200003, Israel}
\affiliation{The Helen Diller Quantum Center, Technion, Haifa, 3200003, Israel}

\begin{abstract}
We study the dynamical spin and nematic correlations in the bilinear-biquadratic spin-1 chain in the critical phase hosting deconfined spinons.
We demonstrate how spinon interactions can be directly probed
in the presence of a magnetic field or a single-ion anisotropy.
Our analytical predictions are supported by numerical matrix-product-state (MPS) simulations of the underlying microscopic model.
\end{abstract}

\maketitle
\section{Introduction}
\label{sec: intro}
The antiferromagnetic bilinear-biquadratic (BLBQ) spin-1 chain is a paradigmatic model in the study of topological phases and fractionalized excitations. As the relative strength between the bilinear and biquadratic interactions is varied, the model exhibits several quantum phase transitions.
At the purely bilinear antiferromagnetic point, the Hamiltonian reduces to the antiferromagnetic Heisenberg model, which for integer spins hosts a gapped, symmetry-protected topological (SPT) ground state known as the Haldane phase~\cite{haldane1983continuum,pollmann2012SPT}.  
The Haldane phase remains stable up to the Uimin-Lai-Sutherland (ULS) point, where the bilinear and biquadratic couplings are equal in strength. At this point, the system undergoes a quantum phase transition into an extended critical phase that persists up to the purely biquadratic limit~\cite{uimin1970one,lai1974lattice,sutherland1975model,itoi1997extended,Lauchli2006,VorosPenc2021}. This phase transition is accompanied by the deconfinement of fractionalized spinon quasiparticles.

The low energy physics in the 
vicinity of the phase transition is captured by the SU(3)$_1$ Wess-Zumino-Witten conformal field theory (CFT) augmented by two marginal interactions: an SU(3)-symmetric backscattering interaction, akin to the marginal backscattering interaction in the low energy description of the spin-1/2 Heisenberg chain, and an SU(3)-breaking pairing interaction responsible for the opening of the Haldane gap~\cite{itoi1997extended}. In the critical phase both of these couplings remain marginally irrelevant.

In this work, we study the BLBQ Hamiltonian in the critical regime subjected to an external Zeeman field or an internal single-ion anisotropy. Previous studies of the BLBQ model in presence of these perturbations focused predominantly on its phase diagram~\cite{Parkinson1989,Kiwata1995,Fath1998,blbq_quadratic_Zeeman,Schmitt1996,manmana2011PairUnbinding,Zvyagin2022}.
Here we focus on the resulting spinon dynamics, as accessed via dynamical spin and nematic correlations. We note that while some properties of the dynamical correlations of the spin-1 chain in presence of Zeeman field were discussed in Ref.~\cite{FengTrivedi2022}, the crucial role of interactions between fractionalized excitations of this rich model has not been addressed previously.
Specifically, here we show that dynamical correlations allow probing the strength of the interactions between spinons.

A closely related scenario was previously demonstrated for the case of the spin-1/2 chain, where the marginally irrelevant backscattering coupling was shown to leave a detectable imprint on the transverse dynamical susceptibility in presence of a longitudinal Zeeman field~\cite{keselman2020dynamicalsignatures,povarov2022electron}.
The underlying mechanism is that in presence of an external (or an internal) field the low-momentum response shifts to finite energies, and although marginally irrelevant couplings flow to zero under the renormalization-group (RG) process, they remain finite at nonzero energies and can therefore manifest in dynamical response functions. 

In the present case of a spin-1 system the situation is richer due to the presence of two distinct marginal interactions in the low-energy theory. 
We show that in a magnetic field, the $k=0$ quadrupolar response directly probes the pairing interaction, while long-wavelength dynamical spin correlations reveal the interplay between the two couplings. In the presence of single-ion anisotropy, the pairing interaction can be detected from the $k=0$ spin response.

To study the dynamical correlations within the low-energy theory we use the hydrodynamic approach developed in Ref.~\cite{wang2022hydrodynamics} generalizing it to the case of spin-1 systems. 
We corroborate our analytical predictions using MPS-based numerical simulations of the microscopic spin-1 BLBQ model.

The rest of the paper is organized as follows.
In Sec.~\ref{sec:model} we present the model and review the low-energy picture. In Sec.~\ref{sec:dynamical} we discuss the dynamical spin and quadrupolar susceptibilites. In Sec.~\ref{sec:hydro} we derive the structure of the dynamical correlations in presence of a generic perturbation to the low-energy model analytically using the hydrodynamic approach. In Sec.~\ref{sec:numerics} we outline the MPS-based calculations used to obtain the correlations numerically for the microscopic model. 
In Secs.~\ref{sec:B-field} and \ref{sec:single-ion} we present results for the dynamical correlations in presence of an external magnetic field and single-ion anisotropy respectively. 
In Sec.~\ref{sec:discussion} we summarize and give an outlook of our results. Appendices~\ref{sec:noninteracting physics}-\ref{sec:combining fields} contain further details on the derivations presented and additional numerical results.

\section{Model and Low-Energy Physics}\label{sec:model}

The BLBQ spin-1 chain is described by the Hamiltonian
\begin{equation}
\mathcal{H} = |J|\sum_i\left[\cos(\theta)(\hat{S}_i\cdot \hat{S}_{i+1}) + \sin(\theta)(\hat{S}_i\cdot \hat{S}_{i+1})^2\right].
\end{equation}
The point $\theta_{ULS}=\frac{\pi}{4}$ is known as the ULS point, and the critical phase is stable in the regime $\theta \in [\frac{\pi}{4},\frac{\pi}{2})$. 

In this phase, the system flows to a quantum critical fixed point, and the spinons form a Tomonaga--Luttinger liquid. 
The low-energy effective theory is formulated in terms of two fields, $\psi_L$ and $\psi_R$, representing left- and right-moving spinons in the vicinity of the Fermi points $\pm k_F = \pm \pi/3$. The SU(3) currents for the two sectors $r=R,L$ are given by
\begin{equation}
J_{r}^a=\psi_{r\alpha}^\dagger T^a_{\alpha\beta}\psi_{r\beta},
\end{equation}
with $T^a_{\alpha\beta}=\frac{1}{2}\lambda^a_{\alpha\beta}$~\cite{itoi1997extended}, and $\lambda_{\alpha\beta}^a$ are the Gell-Mann SU(3) generators   
\footnote{For the explicit matrix form of the generators $T^a_{\alpha\beta}$ used
see Appendix \ref{sec: basis transformation to spin-quadrupole}}. 

The low energy effective Hamiltonian takes the form (repeated indices are summed over) 
\begin{align}
\mathcal{H} &= \mathcal{H}_0+\mathcal{H}_1+\mathcal{H}_2,  \\
\mathcal{H}_0 &= v\int dx\,(\psi_R^\dagger(-i\partial_x)\psi_R+ \psi_L^\dagger(i\partial_x)\psi_L), \\
\mathcal{H}_1 &= g_{1}\int dx\,J^a_{L}J^a_{R}, \\
\mathcal{H}_2 &= g_{2}\int dx\, (\sum_{\alpha\beta}2T_{\alpha\beta}^{a}T_{\alpha\beta}^{b})J_{L}^{a}J_{R}^{b}.
\label{eq: low-energy Hamiltonian}
\end{align}
Here, $\mathcal{H}_0$ is the noninteracting part, $\mathcal{H}_{1}$ is an SU(3)-symmetric backscattering interaction, and $\mathcal{H}_{2}$ is an SU(3)-breaking pairing term. 
Bethe ansatz results imply that $g_1<0$,  while the sign of $g_2$ determines the RG flow of the system~\cite{itoi1997extended}.
For $\pi/4 < \theta < \pi/2$, $g_2$ is positive and the system flows to a gapless fixed point with $g_1^*=g_2^*=0$. 
As $\theta$ is decreased below $\pi/4$, the sign of 
$g_2$ changes and the system undergoes a Berezinskii-Kosterlitz-Thouless (BKT) type quantum phase transition to the gapped (Haldane) phase.
At the critical point, $\theta_{ULS}=\pi/4$, $g_2$ vanishes and the system is SU(3)-symmetric.
\section{Dynamical Susceptibility}\label{sec:dynamical}
The spinon spectrum can be probed using the dynamical structure factor. Specifically, we will consider the transverse spin 
\begin{equation}\label{eq:Spm}
    S^{+-}(k,\omega) = \sum _x e^{-ikx}\int _{-\infty}^\infty dt\,e^{i\omega t} \langle S^+(x,t)S^-(0,0)\rangle,
\end{equation}
and quadrupole 
\begin{equation}\label{eq:Qpm}
    Q^{+-}(k,\omega) = \sum _x e^{-ikx}\int _{-\infty}^\infty dt\,e^{i\omega t} \langle(S^+)^2(x,t) (S^-)^2(0,0)\rangle 
\end{equation}
correlations. 
At zero temperature, these are given by 
\begin{equation}
S^{+-}(k,\omega)= \sum_{m} |\langle m|S^-_k|0\rangle|^2\delta(\omega-E_m)
\label{spin DSF formula}
\end{equation}
and
\begin{equation}
Q^{+-}(k,\omega)= \sum_{m} |\langle m|(S^-_k)^2|0\rangle|^2\delta(\omega-E_m)
\label{quad DSF formula}
\end{equation}
respectively, where $|0\rangle$ denotes the ground state of the system and $|m\rangle$ runs over the excited states.
The excitations created by $S^-$ (or $(S^-)^2$), probed by the dynamical correlations, correspond to particle-hole excitations in terms of the spinons, exhibiting sharp linearly dispersing branches at small momenta and a continuum at large momenta.
In the critical phase and in the absence of perturbations, the couplings $g_{1,2}$ flow to zero and the system is described by three non-interacting spinon bands. It is in fact instructive to examine the dynamical correlations in the non-interacting limit, as it gives a clear understanding of the general structure of the expected dynamical correlations, and highlights which features in the response arise from interactions. This analysis is presented in Appendix~\ref{sec:noninteracting physics}. Below, we discuss several features which carry over to the interacting case in a straighforward manner.

The spectrum of $S^{+-}(k,\omega)$ at zero field features three soft modes, at $k=0$ and $\pm 2k_F$.
When a magnetic field is applied, spinon bands split in accordance with the spin quantum number of a given band. We denote the respective Fermi momenta for the three bands by $k_{F,\alpha}$, with $\alpha=-1,0,1$. For small $B$, the Fermi momenta of the $\alpha=\pm1$ bands shift to $k_{F,\pm1}=k_{F}\pm B/v$, 
while for the $\alpha=0$ band it remains unchanged $k_{F,0}=k_{F}$.
As a result, the two modes at $\pm 2k_F$ each bifurcate into two, yielding four soft modes at $\pm(k_{F,1}+k_{F,0}),\pm(k_{F,0}+k_{F,-1})$, or approximately $\pm 2k_F \pm B/v$ for small $B$~\cite{FengTrivedi2022}. The $k=0$ mode shifts to energy $B$, as the Larmor theorem predicts for SU(2)-symmetric systems in the presence of a magnetic field~\cite{oshikawa2002LarmorESR}.

For the spectrum of $Q^{+-}(k,\omega)$, the contribution comes from scatterings between the $\pm1$ bands. There are two soft modes at $\pm(k_{F,1}+k_{F,-1})$. The generalized Larmor response $\omega(k=0)=2B$ is no longer protected by SU(2) symmetry. Rather, it is protected by the enlarged SU(3)  symmetry, present only in the ULS point for $J_{1,2}>0$ (see Appendix~\ref{sec:quadrupole Larmor}).   

In the following, we analyze how the couplings $g_{1,2}$ affect the dynamical structure factors in the presence of Zeeman field or single-ion perturbations. As mentioned above, these perturbations shift the small-momentum response to finite energies, where the marginally irrelevant couplings remain finite and thus affect the spectrum. For spin-$\frac{1}{2}$ systems, the backscattering interaction was shown to open a spectral gap at finite energies at $k=0$~\cite{keselman2020dynamicalsignatures}. We expect a similar gap opening at $k=0$ for the spin-$1$ model, resulting from hybridization of degenerate modes. The additional pairing term $g_2$ breaks SU(3) symmetry, and its effect on the spectrum is of primary interest in this work.
\subsection{Hydrodynamic Approach}\label{sec:hydro}

Following the hydrodynamic approach introduced in  Ref.~\cite{wang2022hydrodynamics}, we start by deriving the Heisenberg equations of motion for the SU(3) currents using the Kac-Moody commutation relations~\cite{gogolin2004bosonization}
\begin{multline}
    \left[ J^a_{r}(x),  J^b_{r'}(x') \right] = \\ 
    \delta_{r,r'} \left[ \frac{-ir}{4\pi}\delta^{ab}\delta'(x-x')+i\delta(x-x')f^{abc}J_{r}^c(x) \right],
    \label{eq: Kac-Moody}
\end{multline}
where $f^{abc}$ are the SU(3) structure constants.

To simplify the pairing interaction term in Eq.~\eqref{eq: low-energy Hamiltonian}, we use the following relation for the Gell-Mann generators,
\begin{equation}
    2\sum_{\alpha\beta}T^a_{\alpha\beta}T^b_{\alpha\beta} = 
    \begin{cases}
-\delta^{ab}, & a=b=2,5,7 \\
\delta^{ab}, & a=b=1,3,4,6,8 .
\end{cases}
\label{g2 generators mult term}
\end{equation}
This relation can be obtained by noting that $\sum_{\alpha\beta}T^a_{\alpha\beta}T^b_{\alpha\beta} = {\rm Tr}\left[\right(T^a\left)^{\rm T} T^b\right]$, and using $(T^a)^{\rm T} = \mp T^a$ for $a\in\{2,5,7\}$ and $\{1,3,4,6,8\}$ respectively.
For brevity, we denote $\zeta_a = 2\sum_{\alpha\beta}T^a_{\alpha\beta}T^a_{\alpha\beta}$, and introduce 
\begin{equation}
    g_a = g_1+\zeta_a g_2 =
    \begin{cases}
g_1-g_2, & a=2,5,7 \\
g_1+g_2, & a=1,3,4,6,8.
\end{cases}
\label{ga}
\end{equation}
Henceforth, we denote $g_\pm=g_1\pm g_2$.

We derive the Heisenberg equations of motion for the currents in the Gell-Mann basis. To calculate the spin and quadrupolar dynamical correlations, we subsequently transform to the spin-quadrupole basis. This basis consists of three spin operators and five quadrupole operators,
\begin{align}
& Q_1 \equiv (S^x)^2-(S^y)^2, \quad Q_2 \equiv \{S^x,S^y\}, \quad Q_3 \equiv \{S^x,S^z\}, \nonumber \\
& \quad Q_4\equiv \{S^y,S^z\}, \quad Q_5 \equiv (S^x)^2+(S^y)^2-2(S^z)^2, \label{eq:quadrupole operators definition}
\end{align}
where \{.,.\} denotes the anticommutator.  The transformation between these bases is given in Appendix \ref{sec: basis transformation to spin-quadrupole}, where we show that the spin operators are spanned by $\{T^a|a=2,5,7\}$ and the quadrupole operators are spanned by $\{T^a|a=1,3,4,6,8\}$. Equation~\eqref{g2 generators mult term} then implies that the current-current interactions in Eq.~\eqref{eq: low-energy Hamiltonian} separate into decoupled spin and quadrupolar sectors. 

The magnetization and current operators are defined by
\begin{equation}
M^a=J_R^a+J_L^a,\quad 
J^a=J_R^a-J_L^a,
\label{magnetization-current definition}
\end{equation}
where the index $a$ runs over all values from $1$ to $8$.

We consider a general perturbation described by the Hamiltonian
\begin{equation}
    \mathcal{H}_B=-\int dx\,\left[ \sum_aB_a M^a(x) \right].
\end{equation}
Under this perturbation, Heisenberg equations of motion for the magnetizations and the currents are given by
\begin{align}
\partial_t M^a\! &=\! \left(-v+\frac{g_a}{4\pi}\right)\partial_xJ^a\! +\! f^{bac}B^bM^c\! -\! (g_b-g_c)f^{bac}J_L^bJ_R^c, \nonumber \\
\partial_tJ^a\! &=\! \left(-v-\frac{g_a}{4\pi}\right)\partial_xM^a\! +\! f^{bac}B^bJ^c\! -\! (g_b+g_c)f^{bac} J_L^bJ_R^c.
\label{eq:eoms}
\end{align}

We next expand the magnetization and the current around their mean-field values assuming small fluctuations,
\begin{equation}
M^a =m^a+\delta m^a(x,t),\quad
J^a =j^a+\delta j^a(x,t).
\end{equation}
Substituting the mean-field solution into the interaction term yields
\begin{equation}
\mathcal{H}_1+\mathcal{H}_2 \approx \int dx\,\sum_a g_a(J_L^aj_R^a+j_L^aJ_R^a) ,    
\end{equation}
which, together with the perturbation term, gives
\begin{equation}
m^a =  \frac{\chi_0B^a}{1+\frac{1}{2}\chi_0g_a} = \frac{\chi_0 B^a}{1+\delta_a},\quad
j^a = 0,
\end{equation} 
where $\chi_0 = 1/(2\pi v)$ is the free spinon susceptibility, and we have defined $\delta_a = g_a/(4\pi v)$.

To first order in the fluctuations,
\begin{multline}
    (g_b\pm g_c)f^{bac}J_L^bJ_R^c \approx \\ \frac{1}{2}(g_b\pm g_c)f^{bac}m^b\delta j_R^c + \frac{1}{2}(g_b\pm g_c)f^{bac}m^c\delta j_L^{b}, \label{eq: quadratic term}
\end{multline}
where a summation over the indices $b,c$ is implied.
Substituting Eq.~\eqref{eq: quadratic term} into Eq.~\eqref{eq:eoms}, we obtain a system of linear equations in $\delta m^a,\delta j^a$.
(For more details see Appendix~\ref{sec: equations of motion auxiliary data}.)
Introducing the shorthand notation for the vector of magnetization and current fluctuations, $\boldsymbol{\delta\psi}=(\delta m^a, \delta j^a)$ the linearized equations of motion take the form
\begin{equation}
    \partial_t \boldsymbol{\delta\psi}=\mathcal{A}\boldsymbol{\delta\psi}.
\end{equation}
Here $\mathcal{A}$ is a matrix containing numbers and first derivatives in space. After Fourier transformation, we obtain 
\begin{equation}
\omega\boldsymbol{\delta\psi}=\mathcal{A}(k)\boldsymbol{\delta\psi}.
\label{linear eq system form}
\end{equation}

At zero temperature, the dynamical structure factor is proportional to the imaginary part of the Green's function,
\begin{equation}
    \mathcal{G}^{ab}(x,t;x',t')=-i\theta(t-t')\left\langle\left[\delta\psi^a(x,t),\delta\psi^b(x',t')\right]\right\rangle,
\end{equation}
which, as detailed in Appendix~\ref{sec: green function derivation from hydrodynamic equations}, can be calculated directly from Eq.~\ref{linear eq system form}
\begin{equation}
    \mathcal{G}=(\omega - i\mathcal{A}(k) + i0^+)^{-1} \mathcal{F}(k),
    \label{green function expression}
\end{equation}
with 
\begin{equation}
    \mathcal{F}(k)=\mathcal{F}_x\left\langle\left[\delta \psi^a(x,t=0), \delta \psi ^b(0,t=0)\right]\right\rangle.
    \label{F matrix definition}
\end{equation}

\subsection{Numerical Simulation}\label{sec:numerics}

To test our analytical predictions, we compute the dynamical correlations numerically using MPS-based techniques.
Our analysis is carried out on finite chains with $N=200$ sites and open boundary conditions.
We first compute the ground state of the system using the density matrix renormalization group (DMRG) algorithm~\cite{white1992dmrg,Schollwoeck2011}. To obtain the time- and space-resolved correlations we quench the state by applying $S^-$ (or $(S^-)^2$) in the middle of the chain, and simulate the dynamics using the time-evolving block decimation (TEBD) algorithm~\cite{Vidal2004,Paeckel2019}, employing a fourth-order Suzuki-Trotter decomposition. 
We use a time step of $0.05J_1^{-1}$ and evolve the system up to a total time of $T=32\,J_1^{-1}$ (before the light cone of the excitation reaches the boundary). We compute the correlations between the site undergoing the initial quench and the right half of the system and carry out symmetrization with respect to the center of the chain. 
The dynamical spin and quadrupole structure factors, Eqs.~\eqref{eq:Spm},\eqref{eq:Qpm}, are then obtained by a spatial Fourier transform followed by an inverse Fourier transform in time. To enhance the frequency resolution and suppress ringing artifacts arising from the finite simulation time, we apply linear prediction~\cite{white2008linearprediction} together with a Gaussian windowing function, $w(t)=\exp(-2t^2/T_{\rm ext}^2)$. For the magnetic field simulations, we extrapolate the data to $T_{\rm ext}=2T$. The single-ion analysis requires higher resolution, and we therefore extrapolate to $T_{\rm ext}=3T$. 
We make sure the extrapolation followed by the windowing does not create new features or significantly alter the branches and only enhances the resolution. 
Throughout the analysis we use bond dimensions up to $M=400$ for which the truncation errors are of order $10^{-6}$.
Numerical calculations were carried out using the ITensor library~\cite{itensor,itensor-r0.3}.

\section{Magnetic Field}\label{sec:B-field}

We start with analyzing the response to an external magnetic field. The coupling to the magnetic field is given by 
\begin{equation}
    \mathcal{H}_B=-B\sum_i S^z_i .
\end{equation}
Below, we incorporate this field into the continuum model, obtain the dynamical structure factors from the linearization of Eq.~\eqref{eq:eoms}, and compare with the numerical results.
\subsection{Analytical Results}

In the continuum the coupling to the field is given by
\begin{equation}
    \mathcal{H}_B=-B\int dx\,M^z=2B\int dx\,M^5(x).
\end{equation}

Introducing 
\begin{equation}
    \eta_{i}=-\frac{\chi_0 g_{i}}{1+ \delta_-}=-\frac{2\delta_{i}}{1+\delta_-},
    \label{eq: eta definition}
\end{equation}
the equations of motion for the fluctuations in the spin sector ($a\in \{ 2,5,7 \} $) are given by
\begin{align}
\partial_t \delta m^a &= -v(1-\delta_-)\partial_x \delta j ^a+f^{bac}B^b \delta m^c, \nonumber \\
\partial_t \delta j^a & =-v(1+\delta _-)\partial_x\delta m^a+(1+\eta_-)f^{bac}B^b\delta j^c.
\label{eq:spin eoms simplified}
\end{align}
In the quadrupole sector ($a\in \{ 1,3,4,6,8\}$) the equations are given by
\begin{align}
\partial_t \delta m^a &= -v(1-\delta_+)\partial_x\delta j^a+(1-\eta_2)f^{bac}B^b\delta m^c, \nonumber \\
\partial_t\delta j^a &= -v(1+\delta_+)\partial_x\delta m^a+(1+\eta_1)f^{bac}B^b\delta j^c.
\label{eq:quadrupole eoms simplified}
\end{align}

The transverse spin dynamical susceptibility, obtained from Eq.~\eqref{green function expression} with $\mathcal{A}(k)$ determined by Eq.~\eqref{eq:spin eoms simplified}, 
\begin{equation}
\chi^{\pm}_S(k,\omega) = \chi_0\left[\frac{A_+^{S}}{\omega-\omega_+^{S}(k) + i0^+}+\frac{A_-^{S}}{\omega-\omega_-^{S}(k) + i0^+}\right] ,
\label{DSF general form}
\end{equation}
where
\begin{align}
\omega_\pm^{S} &= \frac{B}{1+\delta_-} \pm \sqrt{\left(\frac{B\delta_- }{1+\delta_-}\right)^2+\left(1-\delta_-^2\right)(vk)^2}, 
\label{eq:spin resonance} \\
A_\pm^{S} &= 4\left( \frac{B}{1+\delta_-} \pm \frac{\delta_- \left(\frac{B}{1+\delta_-}\right)^2+(1-\delta_-)(vk)^2}{\sqrt{\left(\frac{\delta_-B}{1+\delta_-}\right)^2+(1-\delta_-^2)(vk)^2}}\right).
\label{eq:spin residue}
\end{align}

These results agree with those of Ref.~\cite{keselman2020dynamicalsignatures}, with $\delta_-$ playing the role of the effective normalized backscattering coupling $\delta$ (note the opposite sign convention) in the spin-1/2 case.
In particular, the spectral gap at $k=0$ closes when $\delta_-=\delta_2-\delta_1=0$. Since $\delta_1<0$ throughout the parameter space, the point $\delta_-=0$ lies in the regime $J_2 < J_1$. In addition, for any $\delta_-$ the resonance at $k=0$ is at $\omega=B$, satisfying the Larmor theorem. \\

In Fig.~\ref{fig:Spins Magnetic Field Analytical DSF, delta_2=-0.05} we plot $\omega_\pm^{S},A_\pm^{S}$ at the fine-tuned point corresponding to a vanishing backscattering $\delta_2=\delta_1$, as well as for $\delta_-\neq0$ (specifically, we take $\delta_2>0$, corresponding to $J_2>J_1$). 

We note that in the parameter regime used in the numerical analysis of the microscopic model $|\delta_2|<|\delta_1|$ (see Appendix~\ref{sec: magnetic field additional numerical results}), and therefore $\delta_\pm < 0$.
The spectral gap at $k=0$ can then be written explicitly as
\begin{equation}
    \Delta_{\pm}^{S} =  -2B\frac{\delta_-}{1+\delta_-} = 2B\frac{|\delta_-|}{1+\delta_-}.
    \label{eq: spin spectral gap at $k=0$}
\end{equation}

\begin{figure} 
    \centering
    \includegraphics[width=\columnwidth]{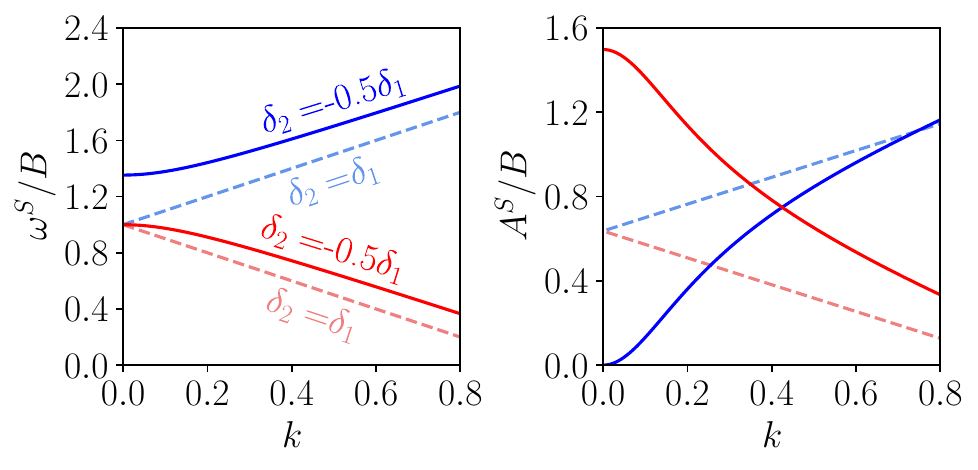}
    \caption{The dispersion, $\omega_{\pm}^{S}$ (see Eq.~\eqref{eq:spin resonance}), and spectral weight, $A_{\pm}^{S}$ (see Eq.~\eqref{eq:spin residue}), of the sharp branches expected in the transverse spin dynamical correlations at small momenta. The plots are given for $B=1$, $v=1$, $\delta_1=-0.1$, with  $\delta_2=\delta_1$, i.e., $\delta_-=0$ (dashed lines), and $|\delta_2|<|\delta_1|$ (solid lines).}
    \label{fig:Spins Magnetic Field Analytical DSF, delta_2=-0.05}
\end{figure}

The transverse component of the quadrupolar dynamical susceptibility, obtained from Eq.~\eqref{green function expression} with $\mathcal{A}(k)$ determined by Eq.~\eqref{eq:quadrupole eoms simplified}, is given by
\begin{equation}
\chi^{\pm}_Q(k,\omega) = \chi_0\left[\frac{A_+^{Q}}{\omega-\omega_+^{Q}(k) + i0^+}+\frac{A_-^{Q}}{\omega-\omega_-^{Q}(k) + i0^+}\right] ,
\end{equation}
where
\begin{align}
\omega_\pm^{Q} &= \frac{2B}{1+\delta_-} \pm \sqrt{\left(\frac{2B\delta_+}{1+\delta_-}\right)^2+\left(1-\delta_+^2\right)(vk)^2}, 
\label{eq:quad resonance} \\
A_\pm^{Q} &= 4\left(\frac{2B}{1+\delta_-}\pm \frac{4B^2\frac{\delta_+}{(1+\delta_-)^2}+(1-\delta_+)(vk)^2}{\sqrt{\left(\frac{2B\delta_+}{1+\delta_-}\right)^2+(vk)^2 \left(1-\delta_+^2\right)}}\right).
\label{eq:quad residue}
\end{align}

In contrast to the spin case, there is no Larmor theorem protecting the quadrupole response at $k=0$. For $g_2=0$, the lower branch exhibits a resonance at $2B$, which corresponds to the $1\to -1$ transition energy of a noninteracting system. This result is exact at the microscopic level for an SU(3) symmetric system coupled to a magnetic field, as we prove in Appendix \ref{sec:quadrupole Larmor}. When $g_2$ becomes finite, SU(3) symmetry is broken and the lower branch resonance at $k=0$  deviates from $2B$. 
The shift from $\omega=2B$ at $k=0$ to lowest order in $\delta_2,\delta_-$ is
\begin{equation}\label{eq: quad shift}
\Delta_{-}^{Q} =  2B\left(\frac{1+\delta_+}{1+\delta_-} - 1\right) \simeq 4B\delta_2 -4B\delta_2(\delta_1-\delta_2).
\end{equation}
The shift is proportional to $\delta_2$, i.e negative for $\delta_2<0$  ($J_2<J_1$), and positive for $\delta_2>0$  ($J_2>J_1$). 

 In Fig.~\ref{fig:Quad Magnetic Field Analytical DSF, delta_1=-0.1} we plot $\omega_\pm^{Q},A_\pm^{Q}$  at the SU(3) symmetric point $\delta_2=0$ corresponding to $J_2=J_1$, as well as for $\delta_2\neq0$ (specifically, we take $\delta_2>0$, corresponding to $J_2>J_1$).
\begin{figure}
    \centering
    \includegraphics[width=\columnwidth]{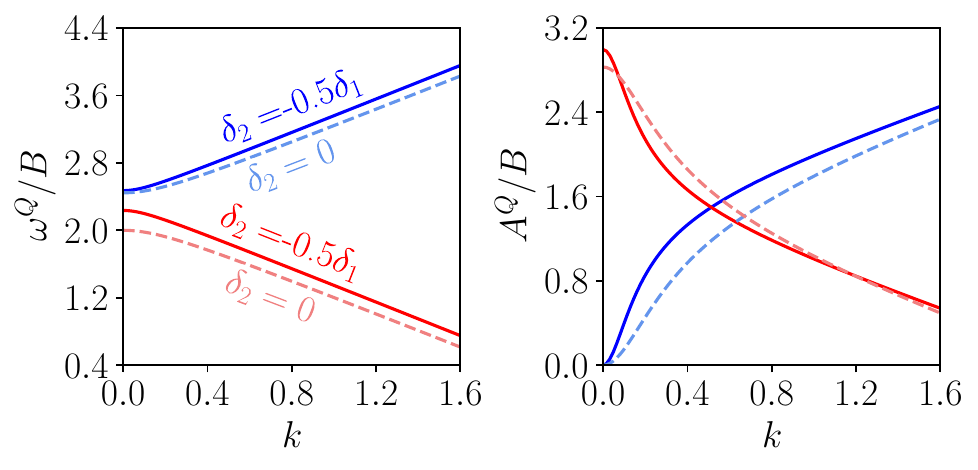}
    \caption{The dispersion, $\omega_{\pm}^{Q}$ (see Eq.~\eqref{eq:quad resonance}), and the spectral weight, $A_{\pm}^{Q}$ (see Eq.~\eqref{eq:quad residue}), for the sharp branches expected in the quadrupole dynamical correlations, for $B=1$, $v=1$, $\delta_1=-0.1$, plotted for $\delta_2=0$ (dashed lines) and $\delta_2>0$ (solid lines).}
    \label{fig:Quad Magnetic Field Analytical DSF, delta_1=-0.1}
\end{figure}

\subsection{Numerical Results}
\label{sec:magnetic field numerical results}

\begin{figure} 
    \centering
    \includegraphics[width=\columnwidth]{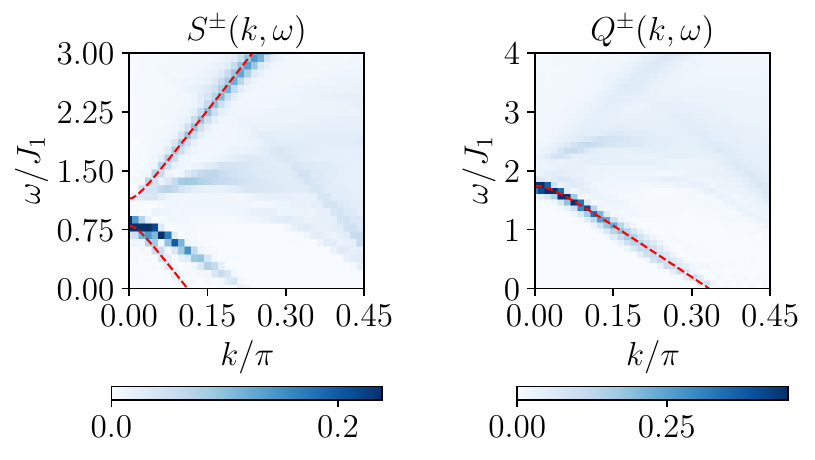}
    \caption{Dynamical spin, $S^{+-}(k,\omega)$, and quadrupolar, $Q^{+-}(k,\omega)$, correlations obtained numerically for the microscopic spin-1 BLBQ model with $B=0.8J_1$, $J_2=1.1J_1$. Dashed lines show fits to the dispersions $\omega_{\pm}^{S}(k)$ (Eq.~\eqref{eq:spin resonance}) and $\omega_-^{Q}(k)$ (Eq.~\eqref{eq:quad resonance}) obtained analytically within the low-energy effective field theory description.}
    \label{fig:DSF for J2=1.1,B=0.8}
\end{figure}

We now turn to the numerical simulations to test our predictions for the resonance branches.
In Fig.~\ref{fig:DSF for J2=1.1,B=0.8} we 
show representative spin and quadrupolar dynamical structure factors obtained for $J_2/J_1=1.1$ and $B/J_1=0.8$.
The presence of four linear branches at small momenta in the spin response is due to the curvature of the bands. When a magnetic field is applied, the Fermi velocities of the different bands change unequally, resulting in different slopes in Fig.~\ref{fig:DSF for J2=1.1,B=0.8}. In our analysis we use the upper upward branch and the lower downward branch in $S^{+-}(k,\omega)$ since these suffer least from band curvature effects as the $-1$ band is depleted (see Appendix~\ref{sec:noninteracting physics}). 
Due to the low intensity of the downward branch we use it only to verify the consistency of the parameters, as discussed below.
In the analysis of the quadrupolar response, $Q^{+-}(k,\omega)$, we use the downward branch which gives direct access to $\delta_2$ through the energy shift at $k=0$ (Eq.~\ref{eq: quad shift}). Since the quadrupole upward branch is broader and has weak spectral weight, we do not include it in our analysis.

Carrying out a systematic analysis, we vary $J_2$, taking it to be in the vicinity of the ULS point, as well as the field strength, $B$, and fit the branches observed numerically to the predicted analytical form. 
(Note that in presence of a magnetic field the critical phase extends to $J_2<J_1$~\cite{Fath1998}.) 
Specifically, for each value of $J_2$ and $B$ we fit the upward branch in the spin response, $\omega^S_+(k)$, and the downward branch in the quadrupole response, $\omega^Q_-(k)$.
These fits allow for an accurate extrapolation of the branches observed numerically to $k=0$, providing access to the gap in the spin response, calculated as $\Delta^S_\pm(B,J_2)=\omega^S_+(k=0)-B$, and the shift of the quadrupole lower-branch, $\Delta^Q_-=\omega^Q_-(k=0)-2B$.
While these fits are performed independently, we verify the consistency of the low energy couplings extracted and confirm their agreement with the downward spin branch~\footnote{
Independent fits of the spin and and quadrupole branches result in somewhat different estimates for $\delta_-$. To verify consistency between these fits, we repeat the fit of the quadrupole branch constraining the value of $\delta_-$ to its value obtained from the fit of the spin branch, letting only $\delta_+$ and $v$ vary. Additionally, to verify consistency with the downward spin branch, we fix $\delta_-$ from the fit of the upward spin branch, and fit only the velocity in the downward branch dispersion. These constrained fits show excellent agreement with the data throughout the considered parameter range.}. 

\begin{figure}
\centering
         \includegraphics[width=0.46\columnwidth]{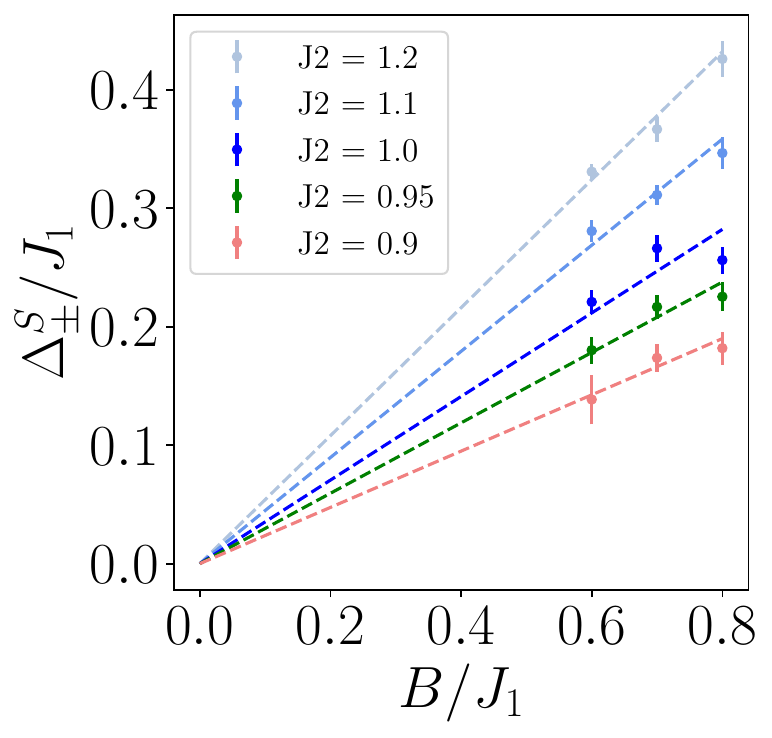}
         \llap{\parbox[c]{0.0in}{\vspace{-3.2in}\hspace{-2.8in}\footnotesize{(a)}}} 
     \hfill
         \includegraphics[width=0.5\columnwidth]{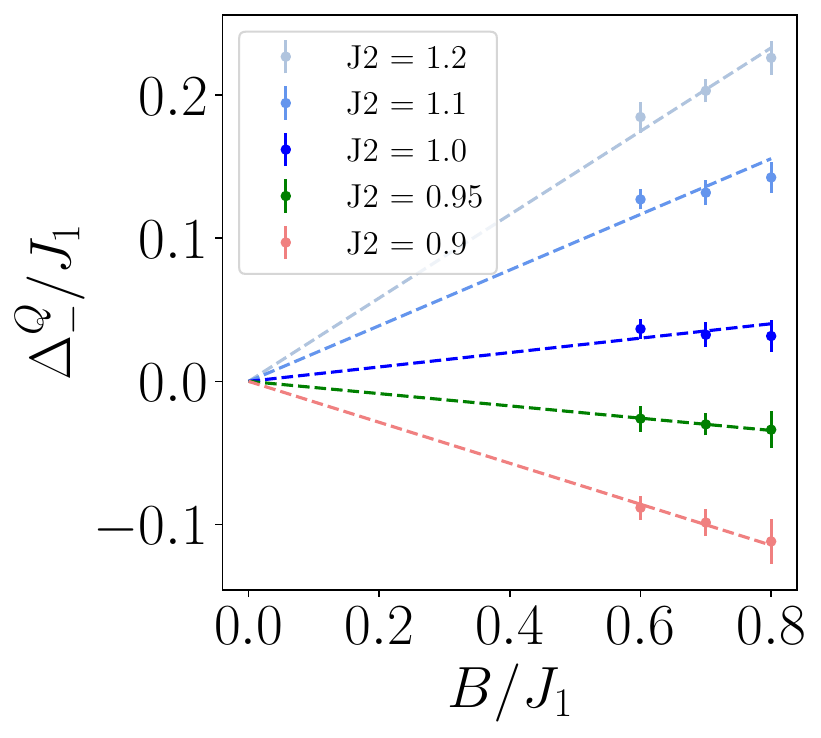}
         \llap{\parbox[c]{0.0in}{\vspace{-3.2in}\hspace{-2.8in}\footnotesize{(b)}}} 
       \caption{(a) The spectral gap at $k=0$ in the spin response, Eq.~\eqref{eq: spin spectral gap at $k=0$}, and (b) the shift in the $k=0$ quadrupole response, Eq.~\eqref{eq: quad shift}, as a function of magnetic field for different values of $J_2$.}
       \label{fig:spin gap (a) and quadrupole lower-branch shift (b) at k=0 vs. magnetic field}
\end{figure}

In Fig.~\ref{fig:spin gap (a) and quadrupole lower-branch shift (b) at k=0 vs. magnetic field} we plot the resulting spin gap, $\Delta^S_\pm$, as well as the quadrupole shift, $\Delta^Q_-$, as functions of the magnetic field, for different values of $J_2$.
For each value of $J_2$, both quantities exhibit a clear linear trend with magnetic field, as expected.
Performing a linear fit of $\Delta^S_\pm$ we extract $\delta_-(J_2)$ using Eq.~\eqref{eq: spin spectral gap at $k=0$}. 
A linear fit of $\Delta^Q_-$ then yields $\delta_+(J_2)$ via Eq.~\eqref{eq: quad shift}, using the value of $\delta_-(J_2)$ extracted beforehand.
The corresponding fits are shown as dashed lines in the figure.
These single-parameter linear fits provide an estimate of the normalized couplings averaged over 
the magnetic fields considered. They agree well with the data, indicating that $\delta_{1,2}$ are indeed nearly constant as a function of $B$ in this regime. 

\begin{figure}
    \centering
\includegraphics[width=0.5\columnwidth]{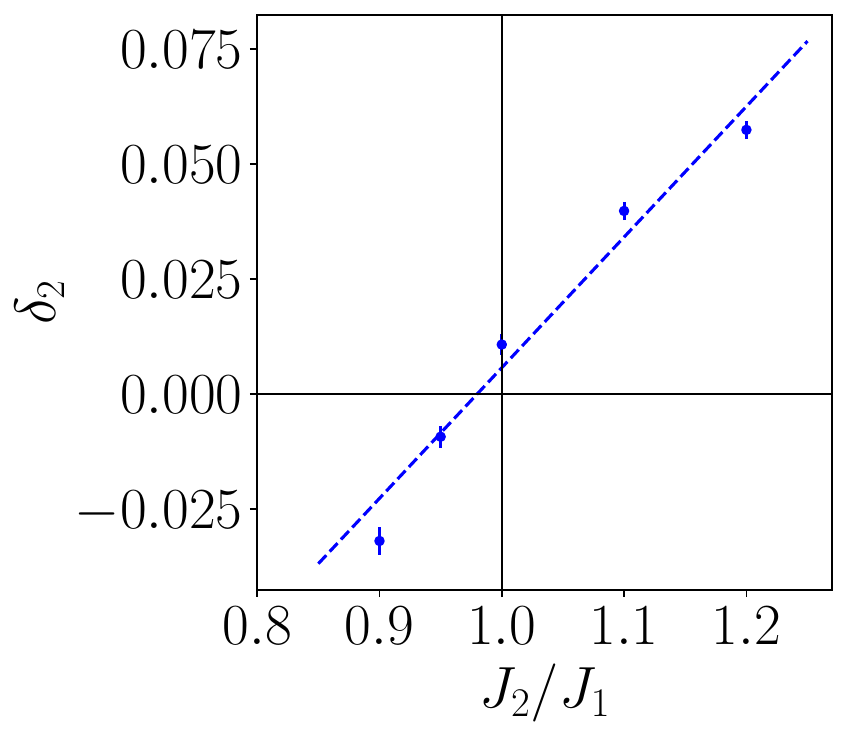}
    \caption{The dimensionless pairing interaction $\delta_2$ as function of $J_2$ obtained from Eq.~\eqref{eq: quad shift} using the data in Fig.~\ref{fig:spin gap (a) and quadrupole lower-branch shift (b) at k=0 vs. magnetic field}.}
    \label{delta_2 vs. J2 for a magnetic field}
\end{figure}

Figure~\ref{delta_2 vs. J2 for a magnetic field} shows the resulting $\delta_2(J_2)$ obtained from the fits above.
As can be seen, $\delta_2$ changes sign with $J_2-J_1$, consistent with the theoretical prediction of Ref.~\cite{itoi1997extended}. 
Our analysis, however, results in a small but finite value of $\delta_2$ at the ULS point, $J_1=J_2$.
This small discrepancy reflects an underestimation of numerical uncertainties, most likely arising from the extraction of the resonance branches.
We note that the vanishing of $\Delta_{-}^{Q}$ at $J_2=J_1$ --- and consequently of $\delta_2$ in our model --- is  an exact result of the microscopic SU(3) symmetry that also applies to finite systems (see Appendix \ref{sec:quadrupole Larmor}).
Hence, we can conclude that the small discrepancy in our results for $\delta_2$ at the ULS point is neither an artifact of the low-energy effective theory nor a finite-size effect in the simulations.

Calculating the dimensionless backscattering coupling, $\delta_1$, we find it to be practically independent of $J_2$ in the parameter range considered, and equal to $\delta_1=-0.15(1)$. 
Additional numerical results presented in Appendix~\ref{sec: magnetic field additional numerical results} confirm that $\delta_1$ does not exhibit a clear trend as a function of $J_2$ in the studied regime.

\section{Single-Ion Anisotropy} \label{sec:single-ion}

In this section we study the effect of a single-ion anisotropy on the dynamical structure factor, following a similar approach to that used in the magnetic field case. 
The single-ion anisotropy is given by 
\begin{equation}
H = -D\sum_i (\hat{S}^{z}_i)^{2},
\end{equation}
where we consider $D>0$.
The single-ion anisotropy splits the $0$ band from the $1,-1$ bands, in analogy with the splitting between the three bands induced by the magnetic field. We may therefore treat both perturbations on equal footing.

\subsection{Analytical Results}
Similar to the magnetic field case, we start by writing the coupling of the single-ion anisotropy to the magnetization operators in the low-energy theory, 
\begin{equation}
\mathcal{H}_D = -D\int (M^3-\frac{1}{\sqrt{3}}M^8)dx\ .
\end{equation}
The equations of motion for the spin sector are given by
\begin{align}
&\partial_t \delta m^a = -v(1-\delta_-)\partial_x \delta j ^a+f^{bac}B^b \delta m^c, \nonumber \\ 
&\partial_t \delta j^a = -v(1+\delta _-)\partial_x\delta m^a+(1+\xi_+)f^{bac}B^b\delta j^c ,
\label{eq:eoms single ion spins}
\end{align}
and for the quadrupole sector
\begin{align}
&\partial_t \delta m^a = -v(1-\delta_+)\partial_x\delta j^a+(1+\xi_2)f^{bac}B^b\delta m^c,  \nonumber \\
&\partial_t\delta j^a=-v(1+\delta_+)\partial_x\delta m^a+(1+\xi_1)f^{bac}B^b\delta j^c,
\label{eq:eoms single ion quadrupoles} \end{align} 
where 
\begin{equation}
    \xi_{i}=-\frac{\chi_0 g_{i}}{1+ \delta_+}=-\frac{2\delta_{i}}{1+\delta_+}.
\end{equation}
The difference between $\xi$ and $\eta$, introduced in Eq.~\ref{eq: eta definition}, is the appearance of $\delta_+$ instead of $\delta_-$ in the denominator. This difference traces back to the spin/quadrupole splitting: the magnetic field couples to the spin sector, while the single-ion anisotropy couples to the quadrupole sector.

Since the single-ion operator $(S^z)^2$ does not split the $1,-1$ bands, we study only the spin dynamical structure factor, which involves hopping between the $0$ band and the $1,-1$ bands. 
The resonance equation for the two branches, $\omega^S_\pm(k)$, is
given in Appendix~\ref{sec: single-ion additional analytical results}. 
The energies of the two branches at $k=0$ are 
\begin{equation}\label{eq:omega_S_k=0}
\omega_{\pm}^S(k=0)= \frac{D}{1+\delta_+}\sqrt{(\delta_-\delta_++1 \mp 2\delta_1)}.
\end{equation}
Recall that $\delta_1<0$, and hence the lower branch resonance at $k=0$ is given by
\begin{align}
\omega_-^S(k=0) &= D\frac{\sqrt{\delta_-\delta_+ +1+2\delta_1}}{1+\delta_+} \nonumber\\
    &\approx D\left( 1-\delta_2+\delta_1\delta_2+\frac{1} {2}\delta_2^2\right) .
\end{align}

At the SU(3)-symmetric point, $\delta_2=0$, we obtain a resonance at $\omega_-^S(k=0)=D$, corresponding to the $1\to0$ transition in a noninteracting system. When $\delta_2$ becomes finite, this energy shifts. We define the lower spin branch shift,
\begin{equation}
    \Delta_-^{S}=\omega_-^S(k=0) - D.
    \label{single-ion lower branch shift}
\end{equation}

In Fig.~\ref{fig:single-ion analytical DSF, delta_1=-0.1} we plot $\omega_\pm^{S}(k)$ at the SU(3) symmetric point $\delta_2=0$ corresponding to $J_2=J_1$, as well as for $\delta_2\neq0$ (specifically, we take $\delta_2>0$, corresponding to $J_2>J_1$).
\begin{figure}
\centering
\includegraphics[width=0.55\columnwidth]{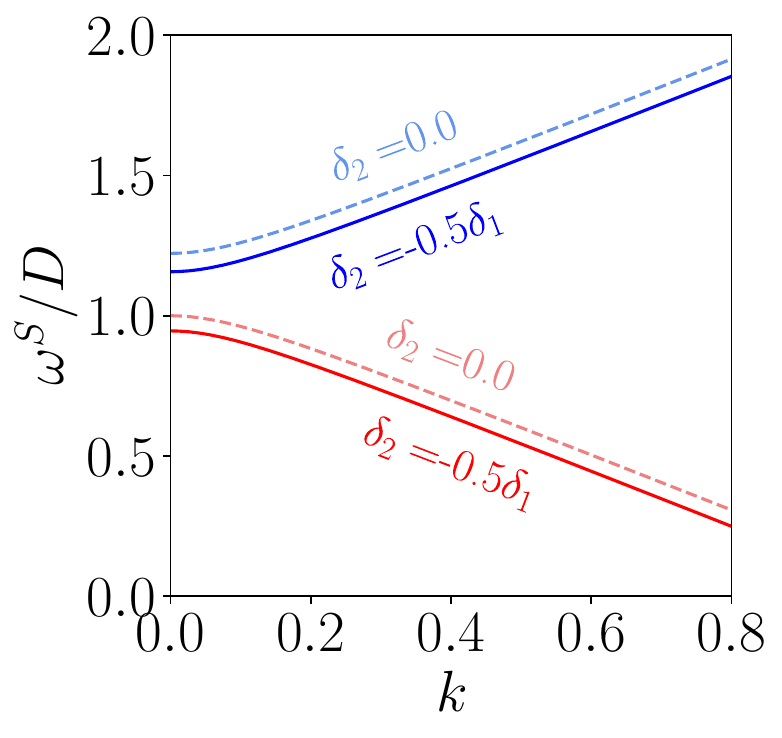}
\caption{The dispersion, $\omega_{\pm}^{S}$, of the branches expected in the spin response in presence of single-ion anisotropy (see Appendix~\ref{sec: single-ion additional analytical results} for the explicit expression). The branches are plotted for $D=1, v=1, \delta_1=-0.1$, with $\delta_2=0$ (dashed lines) and $\delta_2<0$ (solid line). }
\label{fig:single-ion analytical DSF, delta_1=-0.1}
\end{figure}

Obtaining a closed-form expression for the spectral weight in this case is challenging. However, we verify that at $k=0$, the spectral weight of the upper branch vanishes, similar to the magnetic field case and consistent with the numerical results presented below.

\subsection{Numerical Results}
\label{sec: single-ion numerical results}
In contrast to the magnetic field case, in presence of single-ion anisotropy the critical phase cannot be accessed for $J_2<J_1$. In this regime, as $D$ is increased, the system undergoes a transition from the Haldane to the N\'eel-ordered phase~\cite{blbq_quadratic_Zeeman}.
Therefore, in the analysis below, we focus only on the regime $J_2\geq J_1$, where the system is critical, with $\delta_2$ positive and expected to increase with increasing $J_2$.\\

\begin{figure}
\centering
         \includegraphics[width=0.48\columnwidth]{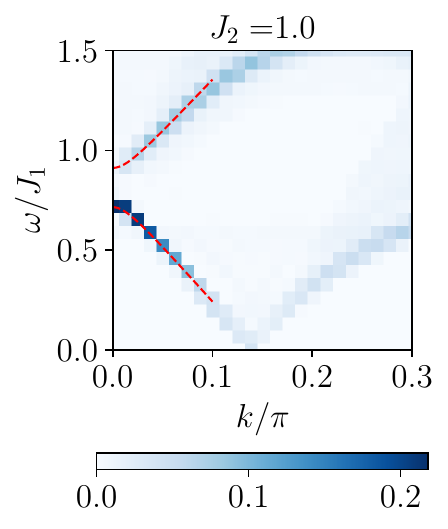}
         \llap{\parbox[c]{0.0in}{\vspace{-3.8in}\hspace{-3.0in}\footnotesize{(a)}}} 
     \hfill
         \includegraphics[width=0.48\columnwidth]{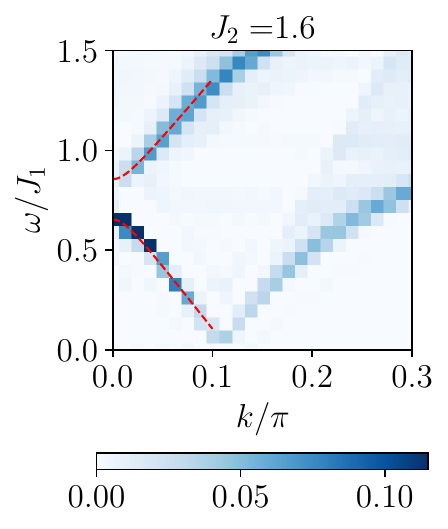}
         \llap{\parbox[c]{0.0in}{\vspace{-3.8in}\hspace{-3.0in}\footnotesize{(b)}}} 
       \caption{Dynamical spin correlations, $S^{+-}(k,\omega)$, obtained numerically for the microscopic spin-1 BLBQ model with single-ion anisotropy $D=0.7J_1$, for (a) $J_2=J_1$ and (b) $J_2=1.6J_1$. Dashed lines show fits to the dispersions obtained within the low energy description, Eq.~\eqref{eq: single-ion branches}.}
       \label{fig: Spin Response D=0.7, J2=1.0,1.6}
\end{figure}

In the case of single-ion anisotropy, our ability to vary the field strength $D$ in the numerical simulations is more limited. For small values of $D$, the frequency resolution is insufficient to extract the couplings with high accuracy. For large $D$, the system moves away from the low-energy regime where the field-theoretical description applies, and the $0$ band becomes depleted. We thus work with a single value of single-ion anisotropy strength, $D/J_1=0.7$, varying $J_2$.

Representative numerical calculations of the dynamical spin response are shown in Fig.~\ref{fig: Spin Response D=0.7, J2=1.0,1.6}. 
To extract $\delta_{1,2}$ for each $J_2$, we first fit the two spin branches to the analytical form expected from the low-energy analysis given in Eq.~\eqref{eq: single-ion branches}. These fits are performed independently and are used for an accurate extrapolation of the branches to $k=0$ similarly to the magnetic field case. Consequently, using $\omega_{\pm}^S(k=0)$, 
we calculate $\delta_{\pm}$ via Eq.~\eqref{eq:omega_S_k=0}.
Once again, to verify consistency of the extracted couplings, we fit both branches again constraining $\delta_\pm$ to the extracted values. We confirm that these fits show very good agreement with the observed dispersions.

\begin{figure}
\centering
\includegraphics[width=\columnwidth]{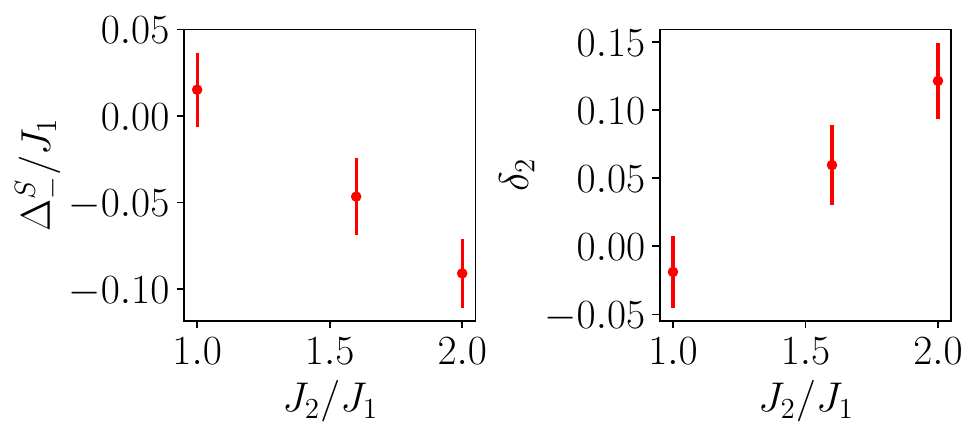}
\caption{$\Delta_-^S$~\eqref{single-ion lower branch shift} and $\delta_2$  (left and right panels, respectively) with single-ion anisotropy $D=0.7J_1$, as a function of $J_2$.}
    \label{fig:D=0.7 couplings}
\end{figure}

In Fig.~\ref{fig:D=0.7 couplings} we show the shift of the lower spin branch, $\Delta_{-}^{S}(J_2)$, defined in Eq.~\eqref{single-ion lower branch shift}, and the extracted dimensionless coupling $\delta_2(J_2)$.

Compared to the magnetic-field case, the dispersion is less sensitive to the coupling strength $\delta_2$, leading to reduced accuracy and larger uncertainties in its extracted value. Nevertheless, we find that $\delta_2$ is close to zero at the ULS point and becomes positive with increasing $J_2$, as expected. Although a wider range of $J_2$ values is explored here, the resulting estimates of $\delta_2$ remain largely consistent with those obtained in the magnetic-field analysis.
The extracted value of $\delta_1$ again shows no explicit dependence on $J_2$ and is found to be equal to $-0.12(3)$, in agreement with the results from the magnetic-field case.

\section{Discussion}\label{sec:discussion}

We studied spinon dynamics in the critical phase of the BLBQ spin-1 chain with two different perturbations - a magnetic field and a single-ion anisotropy. 
Our analytical approach reproduced the structure of the spinon branches at small momenta calculated numerically, allowing us to probe the magnitude of spinon interactions as the microscopic parameters of the spin system are varied.

Specifically, we identified the effect of the SU(3)-breaking pairing interaction on the spectrum. This interaction splits the spin and quadrupole sectors: in presence of a field that couples to one sector, the $k=0$ resonance in the other sector exhibits an energy shift. Strikingly, this shift increases with increasing pairing interaction, $|g_2|$. 
Since in presence of a magnetic field, the critical regime extends to $J_2<J_1$~\cite{Fath1998}, we could probe the response of the system in the critical phase across the ULS point, observing a change in the sign of the energy shift at $J_2=J_1$, consistent with the sign change of $g_2$. We also probed the backscattering coupling $g_1$ and found it to be essentially constant in the studied regime.

While a full analytical treatment in presence of both perturbations remains challenging, we observed that for a weak single-ion anisotropy both the spin and quadrupolar response in presence of external magnetic field is well approximated by the field-theory predictions for the case of a vanishing single-ion anisotropy (see Appendix \ref{sec:combining fields}).

Identifying signatures of spinon interactions in dynamical structure factors opens the route to their experimental observation, as dynamical correlations are directly accessible experimentally.
In particular, spin dynamical structure factor can be measured using
inelastic neutron scattering (INS) ~\cite{zaliznyak2001neutron_scattering_Haldane}.
Probing quadrupolar response is more challenging, but it has recently been demonstrated that it can be accessed via resonant inelastic x-ray scattering (RIXS)~\cite{Haverkort2010,double_spin_flip_probe_RXIS_nature}. In addition, for spin-1 systems, quadrupolar correlations may also be detectable using Raman spectroscopy~\cite{Michaud2011}.

Finally, we note that while finding 1D materials described by the BLBQ spin-1 model with sizable antiferromagnetic $J_2$ remains challenging, optical lattice systems~\cite{garcia2004cold_atoms_BLBQ} with tunable interactions may offer an alternative route to access the critical regime.

\acknowledgements
A.K. and O.A.S. are supported by Grant No. 2024200 from the United States-Israel Binational Science Foundation (BSF).
A.K.\ acknowledges funding by the Israeli Council for Higher Education support program and by the Israel Science Foundation (Grant No.\ 2443/22). 

\appendix

\renewcommand{\thefigure}{A\arabic{figure}}
\setcounter{figure}{0}

\section{Non-interacting Limit}
\label{sec:noninteracting physics}
The presence of three bands in spin-$1$ models enriches the excitation spectrum compared to spin-$\frac{1}{2}$ models, even in the absence of spinon interactions. This Appendix reviews spin and quadrupolar excitations for noninteracting spinon bands in the presence of fields. The noninteracting picture elucidates features in the response of the interacting (spin) model, discussed in the main text.

We consider noninteracting spinons described by the Hamiltonian 
\begin{equation}
\mathcal{H}_0 = -t\sum_{i,\alpha} c_{i,\alpha}^\dagger c_{i+1,\alpha} + h.c.,
\label{free spinons Hamiltonian}
\end{equation}
where $\alpha=-1,0,1$ is the band index (below we use $\bar{1}$ as a shorthand notation for the $-1$ band). 
In the noninteracting limit, the single occupancy constraint only fixes the total density, such that 
\begin{equation}
k_{F,-1}+k_{F,0}+k_{F,1} = \pi,
\end{equation}
where $k_{F,\alpha}$ is the Fermi momentum of band $\alpha$, and we set the lattice constant to unity.

The spin dynamical structure factor, Eq.~\eqref{spin DSF formula}, can be written as
\begin{align}
& S^{+-}(k,\omega)= \sum_{m} |\langle m|S^-_k|0\rangle|^2\delta(\omega-E_m) \nonumber \\
&= \sum_{m,q} |\langle m|c_{k+q,0}^\dagger c_{q,1}+c_{k+q,-1}^\dagger c_{q,0}|0\rangle|^2\delta(\omega-E_m).
\end{align}
Thus, it probes two particle-hole excitation channels: $0\to-1$  and $1\to0$. Similarly, the quadrupolar dynamical structure factor, Eq.~\eqref{quad DSF formula}, can be written as
\begin{equation}
Q^{+-}(k,\omega) = \sum_{m,q} |\langle m|c_{k+q,-1}^\dagger c_{q,1}|0\rangle|^2\delta(\omega-E_m).    
\end{equation}
It probes particle-hole excitations between bands $-1$ and $1$.

We first consider the field-free case, where $k_{F,\alpha}=\pi/3$. The linearized dispersion in the vicinity of the Fermi points results in a sharp linear branch at small momentum, a well known result for particle-hole excitations in 1D. In addition to the soft mode at $k=0$, there are two additional soft modes at $k_s=\pm 2\pi/3$, corresponding to backscattering between opposite Fermi points. These modes sit at the bottom of the two-spinon continua, which broaden with increasing energy.

The magnetic field term is given by,
\begin{equation}
\mathcal{H}_B=-B\sum_i S^z_i = -B\sum_i (c^\dagger_{i,1}c_{i,1} - c^\dagger_{i,-1}c_{i,-1}).
\end{equation}
This term splits the three degenerate spinon bands. 
The single-ion term is given by,
\begin{equation}
\mathcal{H}_D=-D\sum_i (S_i^z)^2=-D\sum_i (c_{i,1}^\dagger c_{i,1} + c_{i,-1}^\dagger c_{i,-1}).
\end{equation}
This term splits the $1,-1$ bands from the $0$ band.

\begin{figure}[t]
    \centering
     \includegraphics[scale=0.17]{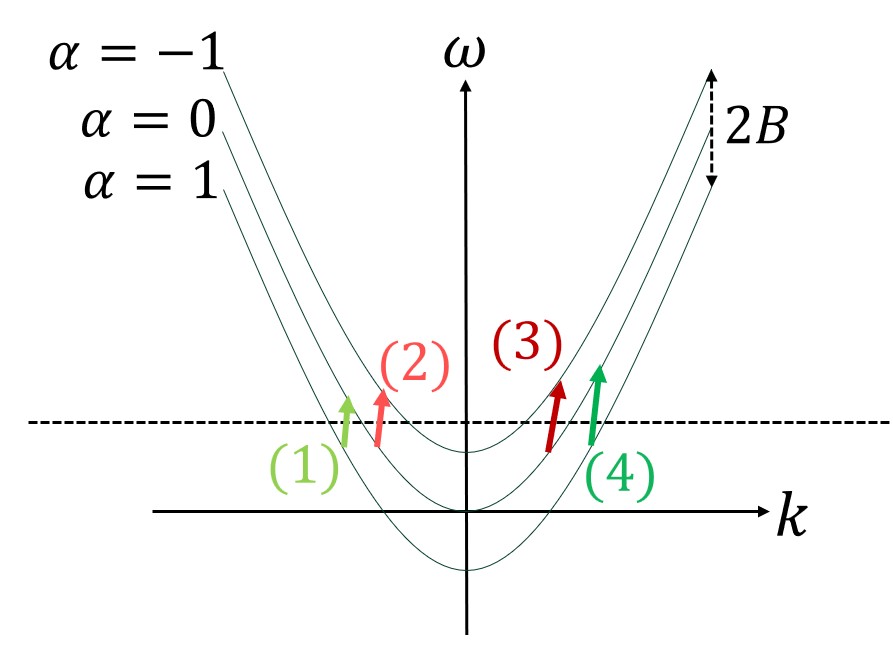}
         \llap{\parbox[c]{0.0in}{\vspace{-2.4in}\hspace{-3in}\footnotesize{(a)}}} 
     \hfill
         \includegraphics[scale=0.17]{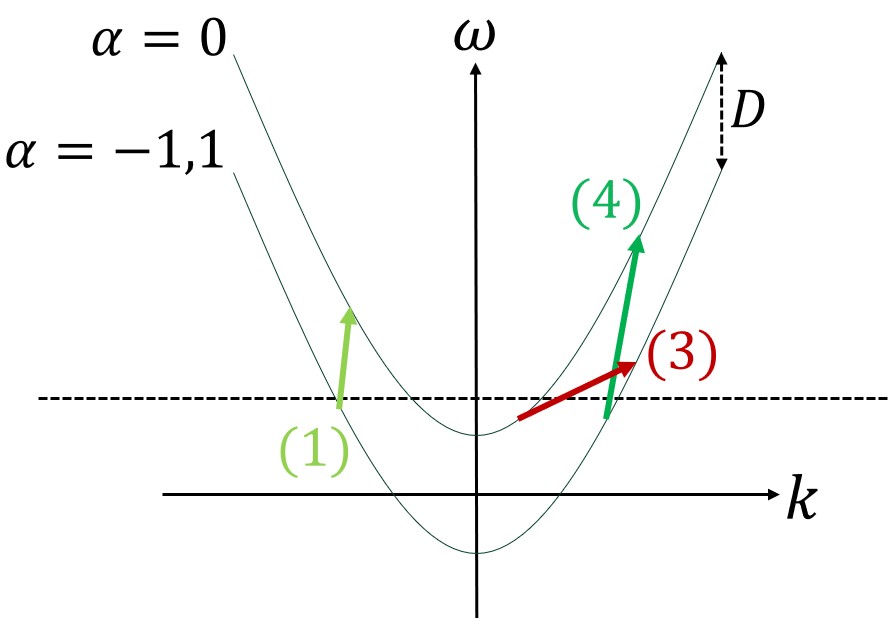}
         \llap{\parbox[c]{0.0in}{\vspace{-2.4in}\hspace{-3in}\footnotesize{(b)}}}
    \caption{Noninteracting spinon bands under (a) magnetic field and (b) single-ion anisotropy. Excitations contributing to the spin dynamical structure factor are sketched with arrows.}
    \label{fig: free bands with fields}
\end{figure}

Figure~\ref{fig: free bands with fields} shows the spinon bands in presence of these two fields. For simplicity, the cosine dispersion is drawn as parabolic. Different particle-hole processes contributing to the structure factors are indicated with arrows.

We focus on the magnetic field case, which is more involved due to the three non-degenerate bands participating in the response. Figure~\ref{fig: Noninteracting spin dynamical structure factor with t=1, for B=1.3 (a) and B=1.5} shows the exact calculation of the spin dynamical structure factor for two different magnetic fields. In this figure we plot the contributions of the two scattering channels, $1\to0$ and $0\to -1$, separately. 
There is a backscattering soft mode in each channel, and its bifurcation from $2\pi/3$ is due to the depletion of the $-1$ band and the increased occupation of the $1$ band~\cite{FengTrivedi2022}. The other soft modes (one for each channel) correspond to a forward scattering of left-moving to left-moving spinons.  As a result of the band curvature, they appear at different momentum in the two channels, with the $1\to 0$ soft mode having a smaller $k$. The downward branch in the $0\to -1$ channel gets wider with decreasing density of $-1$ spinons, as the curvature increases. Finally, it becomes massive when the $-1$ band depopulates.

The different particle-hole branches in Fig.~\ref{fig: Noninteracting spin dynamical structure factor with t=1, for B=1.3 (a) and B=1.5} are shown with labels denoting the scattering processes in Fig.~\ref{fig: free bands with fields}.  
To understand the expected response, take the limit of 
a small field such that the Fermi velocity, $v_F$,
is approximately band-independent,
and consider a scattering $k\to k'$ in the vicinity of one of the Fermi points  ($|k|-k_F,|k'|-k_F\ll k_F$). To linear order, the excitation energy is $B\pm v_F(k'-k)$ 
for the $R,L$ sectors, respectively. Therefore, processes $(1),(2)$ which involve $L\to L$ scatterings correspond to downward branches and processes $(3),(4)$ which involve $R\to R$ scatterings corresponds to upward branches.

The two-spinon continua in Fig.~\ref{fig: Noninteracting spin dynamical structure factor with t=1, for B=1.3 (a) and B=1.5} are denoted by scattering processes $r\alpha\to r'\alpha'$, where $r,r'\in\{R,L\}$ are right- or left-moving spinons,
and $\alpha,\alpha'$ are spinon flavors.

\begin{figure}[t]
\centering
         \includegraphics[scale=0.18]{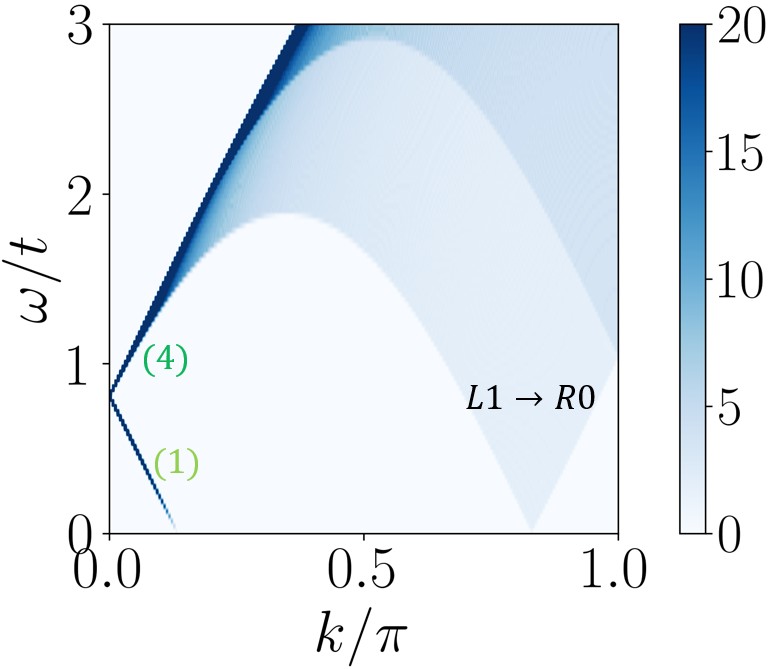}
         \llap{\parbox[c]{0.0in}{\vspace{-2.9in}\hspace{-3.2in}\footnotesize{(a)}}} 
     \hfill
         \includegraphics[scale=0.18]{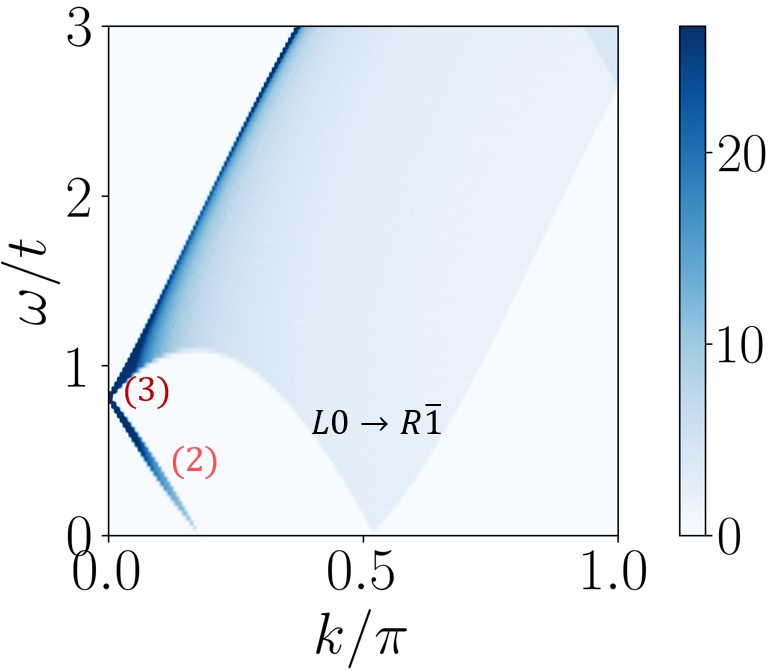}
         \llap{\parbox[c]{0.0in}{\vspace{-2.9in}\hspace{-3.2in}\footnotesize{(b)}}}
         
         \vspace{0.1in}
         
         \includegraphics[scale=0.18]{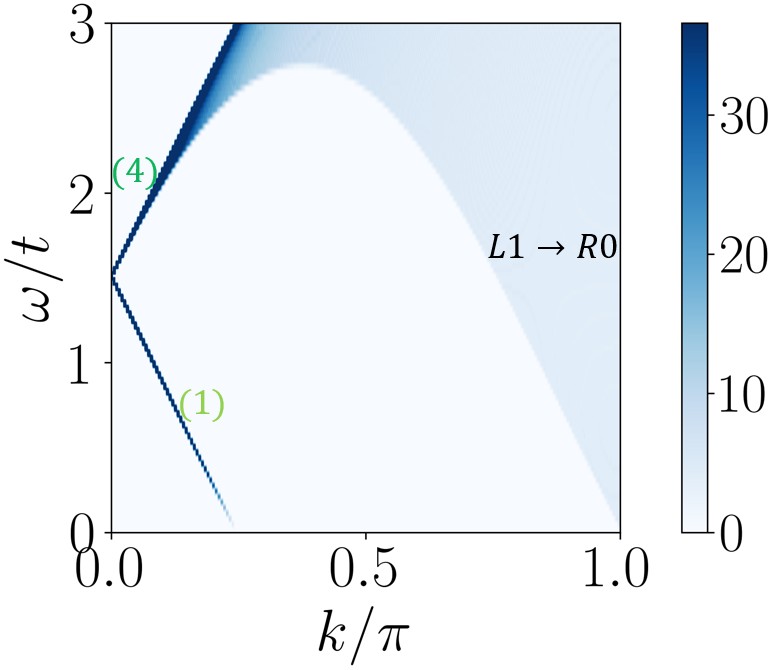}
         \llap{\parbox[c]{0.0in}{\vspace{-2.6in}\hspace{-3.2in}\footnotesize{(c)}}} 
     \hfill
         \includegraphics[scale=0.18]{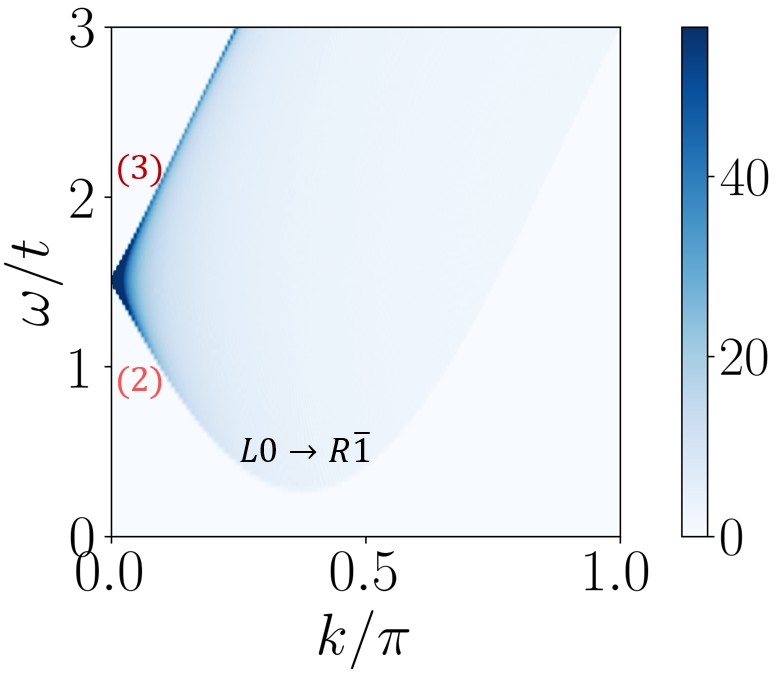}
         \llap{\parbox[c]{0.0in}{\vspace{-2.6in}\hspace{-3.2in}\footnotesize{(d)}}} 
       \caption{Spin dynamical structure factor in the noninteracting limit, in presence of magnetic field, calculated for $B/t=0.8$ in (a,b) and $B/t=1.5$ in (c,d).
       Contributions from the $1\to0$ channel and the $0\to-1$ channel are shown separately in (a,c) and (b,d) respectively. The labels of the different branches follow notations in Fig.~\ref{fig: free bands with fields}.}
       \label{fig: Noninteracting spin dynamical structure factor with t=1, for B=1.3 (a) and B=1.5}
\end{figure}

\begin{figure}[t]
\centering
         \includegraphics[scale=0.2]{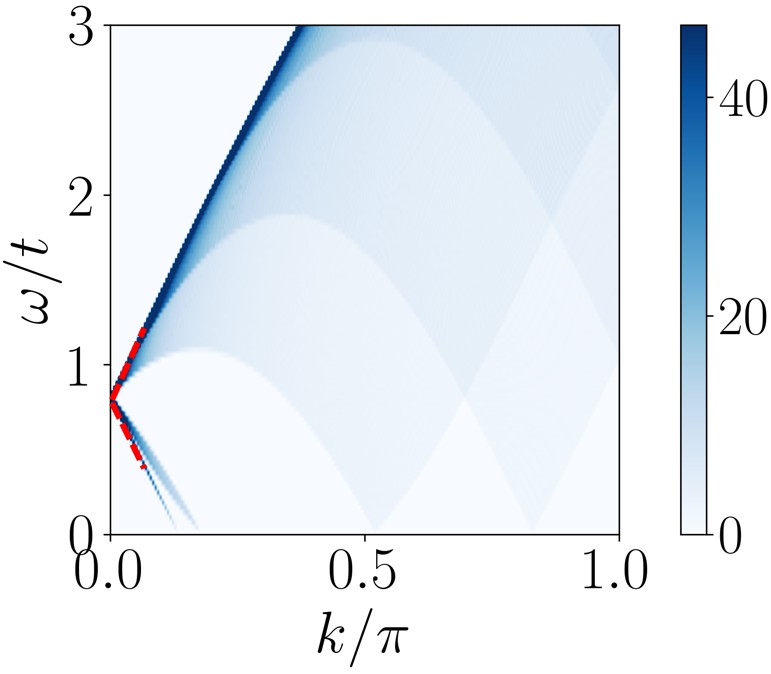}
         \llap{\parbox[c]{0.0in}{\vspace{-2.9in}\hspace{-3.05in}\footnotesize{(a)}}} 
     \hfill
         \includegraphics[scale=0.2]{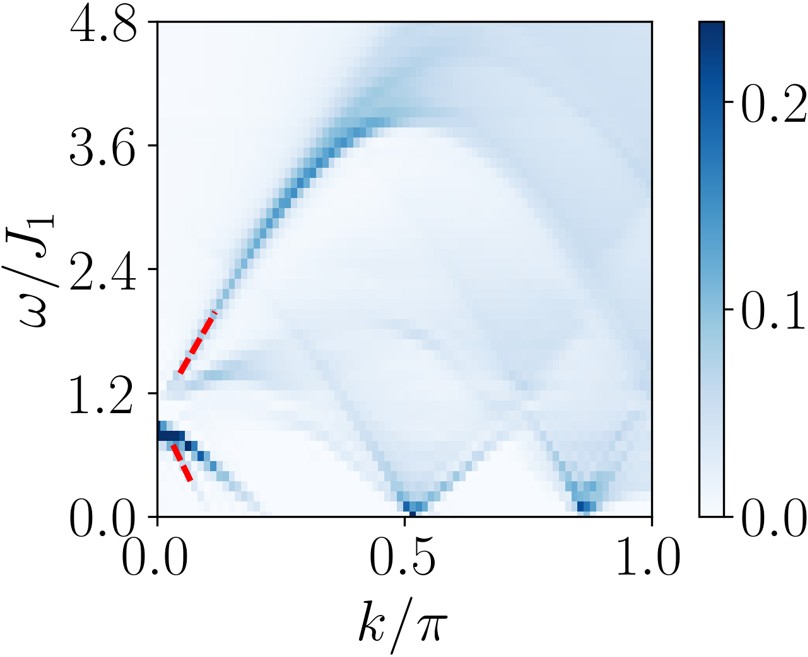}
         \llap{\parbox[c]{0.0in}{\vspace{-2.9in}\hspace{-3.05in}\footnotesize{(b)}}} 
       \caption{Spin dynamical structure factor in magnetic field. (a) Noninteracting response for $B/t=0.8$ and (b) MPS calculation for the spin model for $J_2/J_1=1.1$, $B/J_1=0.8$.}
       \label{fig: noninteracting vs. interacting spin DSF with magnetic field}
\end{figure}

A comparison between the noninteracting response and the MPS calculation is presented in Fig.~\ref{fig: noninteracting vs. interacting spin DSF with magnetic field}. 
In the interacting case, an additional sharp upward branch emerges at small momenta  (below the upward branch marked by dashed red line). We postulate that it appears out of the small-$k$ 
continuum as a result of interactions, which can affect the spectral weight and give rise to singularities~\cite{muller1981quantum}. The red dashed lines in Fig.~\ref{fig: noninteracting vs. interacting spin DSF with magnetic field}(b) highlight the branches used to analyze spinon interactions as discussed in the main text, with the corresponding branches highlighted by red dashed lines also in the noninteracting limit, Fig.~\ref{fig: noninteracting vs. interacting spin DSF with magnetic field}(a).
These branches suffer least from curvature effects, making them suitable for testing the low energy theory.

The two-spinon continua and the bifurcation of the soft modes are also seen in both cases, where again the interactions seem to redistribute the spectral weight and create intensity peaks along the continua boundaries~\cite{muller1981quantum}.

We next discuss the quadrupolar response in presence of a magnetic field. There is a single hopping channel for each mode (upward and downward), and thus we expect to see two linear branches at small k. This is indeed the case, as can be seen for example in Fig.~\ref{fig:DSF for J2=1.1,B=0.8}. The presence of two modes is analogous to spin-$\frac{1}{2}$ systems, where there are only two bands.

\begin{figure}[t]
\centering
    \includegraphics[scale=0.19]{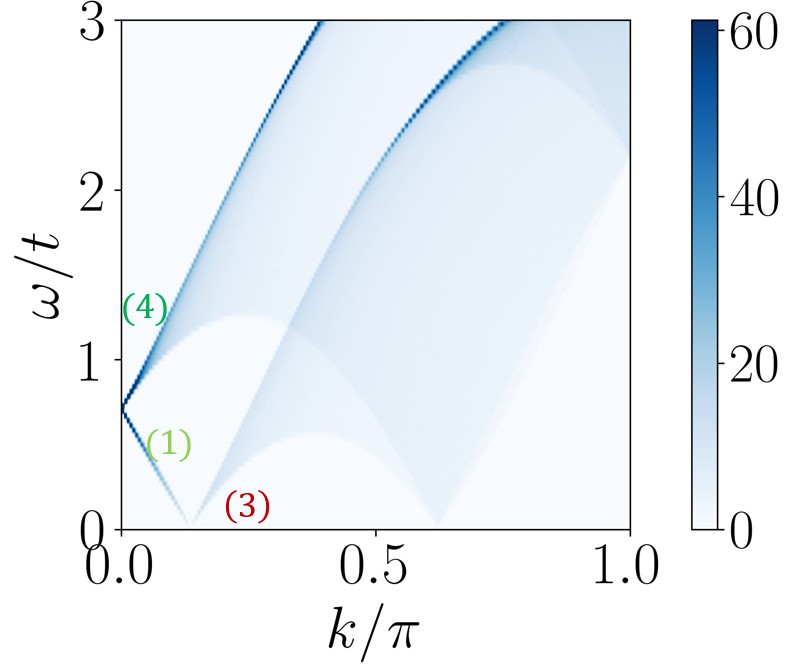}
         \llap{\parbox[c]{0.0in}{\vspace{-2.9in}\hspace{-3.05in}\footnotesize{(a)}}} 
     \hfill
    \includegraphics[scale=0.19]{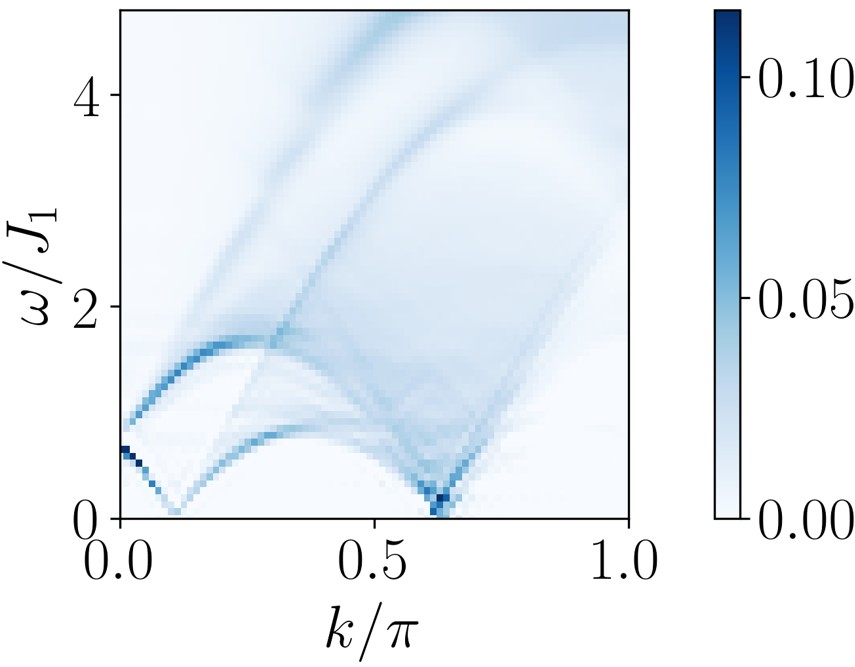}
         \llap{\parbox[c]{0.0in}{\vspace{-2.9in}\hspace{-3.05in}\footnotesize{(b)}}} 
       \caption{Spin dynamical structure factor in presence of single-ion anisotropy. (a) Noninteracting response for $D/t=0.7$ and (b) MPS calculation for $J_2/J_1=1.6$, $D/J_1=0.7$. In (a) the different branches are labeled according to the corresponding processes in Fig.~\ref{fig: free bands with fields}. In the interacting case, hybridization between the modes prevents us from identifying the branches observed with specific inter-band processes.}
       \label{fig: noninteracting vs. interacting spin DSF in single-ion}
\end{figure}

Finally, we discuss the single-ion case. Comparison between noninteracting and spin dynamical structure factors is shown in Fig.~\ref{fig: noninteracting vs. interacting spin DSF in single-ion}. The degeneracy between the $1,-1$ bands results in two branches at small $k$, instead of the four branches which appeared in the magnetic field case (see also Fig.~\ref{fig: Spin Response D=0.7, J2=1.0,1.6}). However, there are still two types of excitations: from the lower to the upper-energy band ($1\to 0$) or vice verse ($0\to -1$). Near $k=0$, the downward and upward branches correspond to $L1\to L0$ and $R1\to R0$, respectively. The downward branch becomes soft at a momentum that we denote by $k^*$  (at small single-ion anisotropy $k^*\approx D/v_F$).
At the same momentum the scattering $R0\to R\bar{1}$ becomes possible, resulting in a V-shape feature, which persists in the interacting case (at least for the $J_2$ values considered in the study). This feature can be seen clearly in Fig.~\ref{fig: Spin Response D=0.7, J2=1.0,1.6}, where two sharp linear modes emanate from 
$k^*\approx0.1\pi$.

\section{Quadrupole Response at $k=0$}
\label{sec:quadrupole Larmor}
In this section, we prove a generalized version of the Larmor theorem~\cite{oshikawa2002LarmorESR} for the quadrupolar response. We show that at the ULS point, $J_2=J_1$, SU(3) symmetry implies that $\omega^Q(k=0)=2B$.

We start by denoting $T^a_{\rm tot}=\sum_i T^a_i$, where $T^a_i$ are the local SU(3) operators.
For an SU(3) symmetric Hamiltonian $\mathcal{H}$, we have    $[T^a_{\rm tot},\mathcal{H}] = 0$.
Since the transformation from the Gell-Mann operators to the spin-quadrupole basis is linear, this in particular implies, $[(S^-)^2_{\rm tot},\mathcal{H}] = 0$.

To understand the response in presence of field, consider $\mathcal{H}_B=\mathcal{H}-BS^z_{\rm tot}$, and calculate the commutator
\begin{equation}
    [\mathcal{H}_B, (S^-)^2_{\rm tot}]=B[-S^z_{\rm tot},(S^-)^2_{\rm tot}] = 2B(S^-)^2_{\rm tot}.
\end{equation}
(Note that due the trivial commutation relations above, the only contribution comes from the field term.)
Acting with this commutator on the ground state, $\left|0\right\rangle$, we obtain 
\begin{equation}
    \mathcal{H}_B(S^-)^2_{\rm tot}\left|0\right\rangle = (E_0+2B)(S^-)^2_{\rm tot}\left|0\right\rangle,
\end{equation}
where $E_0$ is the ground state energy. Namely, that $(S^-)^2_{\rm tot}\left|0\right\rangle$ is an eigenstate of $\mathcal{H}_B$ with energy $E_0+2B$.

Finally, noting that $(S^-)^2_{\rm tot}=(S^-)^2_{k=0}$, and using Eq.~\eqref{eq:Qpm}, it follows that $Q^{+-}(k=0,\omega)\propto\delta(\omega-2B)$. 
\section{Transformation to the Spin-Quadrupole Basis}
\label{sec: basis transformation to spin-quadrupole}

In this Appendix we establish the connection between the representation of the currents in the Gell-Mann basis used in~\cite{itoi1997extended} and the spin-quadrupole basis used to obtain the dynamical susceptibilities.

The BLBQ Hamiltonian can be written as
\begin{align}
&\mathcal{H} = \mathcal{H}_{\rm ULS} + \mathcal{\delta H}, \nonumber \\
&\mathcal{H}_{\rm ULS} = J_1 \sum_i\left[(\hat{S}_i\cdot \hat{S}_{i+1}) + (\hat{S}_i\cdot \hat{S}_{i+1})^2\right], \nonumber\\
&\mathcal{\delta H} = (J_2-J_1) \sum_i (\hat{S}_i\cdot \hat{S}_{i+1})^2,
\end{align}
where $\mathcal{H}_{\rm ULS}$ is SU(3) symmetric, while  $\mathcal{\delta H}$ breaks SU(3) symmetry.
Using the relation between the spin and spinon operators,
\begin{equation}
    \hat{S}_i^a=\sum_{\alpha\beta}c_{i,\alpha}^\dagger S^a_{\alpha\beta} c_{i,\beta},
\end{equation}
$\delta\mathcal{H}$ can be written as a microscopic pairing interaction in the $S^z$ basis,
\begin{align}
& (S_i \! \cdot \! S_j)^2 \! = \! 
 c_{j,1}^\dagger c_{i,\bar{1}}^\dagger c_{i,\bar{1}} c_{j,1} 
    \! + \! c_{j,0}^\dagger c_{i,0}^\dagger c_{i,0} c_{j,0}  
    \! + \! c_{j,\bar{1}}^\dagger c_{i,1}^\dagger c_{i,1} c_{j,\bar{1}} \nonumber \\ 
     &  +\! 
    \left[ c_{j,\bar{1}}^\dagger c_{i,1}^\dagger c_{i,\bar{1}} c_{j,1} 
     \! - \! c_{j,1}^\dagger c_{i,\bar{1}}^\dagger c_{i,0} c_{j,0}
     \! - \! c_{j,\bar{1}}^\dagger c_{i,1}^\dagger c_{i,0} c_{j,0} 
     \! + \! h.c. \right]
\end{align}

In the basis used in Ref.~\cite{itoi1997extended} it is given explicitly as
\begin{equation}
(S_i\cdot S_j)^2  =\sum_{\alpha,\beta=1}^3 c_{i,\alpha}^\dagger c_{j,\alpha}^\dagger c_{j,\beta} c_{i,\beta}.
\end{equation}
The transformation relating the two representations of the biquadratic term is
\begin{equation}\label{eq:basis_transform}
c_{\pm1} = \frac{1}{\sqrt{2}}(-ic_1 \pm c_3),\quad  
c_0 = c_2.
\end{equation}
Note that the $SU(3)$ symmetric interaction,
\begin{equation}
(S_i\cdot S_j) + (S_i\cdot S_j)^2  =\sum_{\alpha,\beta=1,2,3} c_{i,\alpha}^\dagger c_{i,\beta}c_{j,\beta}^\dagger c_{j,\alpha}, 
\end{equation}
is invariant under \eqref{eq:basis_transform}.
This transformation allows us to relate the spin and quadrupole matrices given in the $S^z$ basis to the Gell-Mann SU(3) generators given in the basis corresponding to $c_{1,2,3}$ in Ref.~\cite{itoi1997extended}.

We obtain, for the spin matrices, 
\begin{align}
S^z = \left(\begin{array}{ccc}
0 & 0 & i\\
0 & 0 & 0\\
-i & 0 & 0
\end{array}\right) = -2T^5, \\
S^y = \left(\begin{array}{ccc}
0 & 0 & 0\\
0 & 0 & i\\
0 & -i & 0
\end{array}\right) = -2T^7, \\
S^x = \left(\begin{array}{ccc}
0 & i & 0\\
-i & 0 & 0\\
0 & 0 & 0
\end{array}\right) = -2T^2,
\end{align}
and for the quadrupole matrices,
\begin{align}
Q_1 &= (S^x)^2-(S^y)^2 =\left(\begin{array}{ccc}
1 & 0 & 0\\
0 & 0 & 0\\
0 & 0 & -1
\end{array}\right) = \sqrt{3}T^8+T^3, \\
Q_2 &= S^{x}S^{y}+S^{y}S^{x} =\left(\begin{array}{ccc}
0 & 0 & -1\\
0 & 0 & 0\\
-1 & 0 & 0
\end{array}\right) = -2T^4, \\
Q_3 &= S^{x}S^{z}+S^{z}S^{x} =\left(\begin{array}{ccc}
0 & 0 & 0\\
0 & 0 & 1\\
0 & 1 & 0
\end{array}\right) = 2T^6, \\
Q_4 &= S^{y}S^{z}+S^{z}S^{y} =\left(\begin{array}{ccc}
0 & 1 & 0\\
1 & 0 & 0\\
0 & 0 & 0
\end{array}\right) = 2T^1, \\
Q_5 &= \frac{1}{\sqrt{3}}\left[(S^x)^2+(S^y)^2-2(S^z)^2\right] =\frac{1}{\sqrt{3}}\left(\begin{array}{ccc}
-1 & 0 & 0\\
0 & 2 & 0\\
0 & 0 & -1
\end{array}\right) \nonumber  \\
&= T^8-\sqrt{3}T^3.
\end{align}

\section{Equations of Motion}
\label{sec: equations of motion auxiliary data}
In this appendix, we elaborate on the linearization of the EOMs given in Eq~\eqref{eq:eoms}, leading to Eqs.~\eqref{eq:spin eoms simplified} and~\eqref{eq:quadrupole eoms simplified} in the presence of a magnetic field, and Eqs.~\eqref{eq:eoms single ion spins} and \eqref{eq:eoms single ion quadrupoles} in the presence of single-ion anisotropy.

Firstly, since the mean-field is constant, it vanishes from the derivative terms in the EOMs.
For the field terms, we show that $B^bM^c$ and $B^bJ^c$ have no mean-field contribution
(as expected from self-consistency of the mean-field). For the former we have
\begin{equation}
f^{bac}B^bm^c = f^{bac}B^b\chi_0\frac{B^c}{1+\delta_c}=0,
\end{equation}
due to the antisymmetry of the structure constants. For the latter, this follows trivially, since $j^c=0$.

We now turn to the remaining term in the EOMs, namely the current-current coupling. Dropping terms quadratic in fluctuations we have
\begin{equation}
J_L^bJ_R^c\approx j_L^b j_R^c +j_L^b\delta j_R^c + j_R^c \delta j_L^{b} , 
\end{equation}
where 
\begin{equation}
j_r^b=\frac{\chi_0}{2}\frac{B^b}{1+\delta_b} 
\label{mean field chiral current}
\end{equation}
is the mean-field value of the chiral current.
We consider a field that couples to either spin or quadrupole operators (but not both), which is the case for both perturbations studied in this work.
Since the mean-field current is proportional to the field (Eq.~\eqref{mean field chiral current}), it is non-vanishing in either the spin or quadrupole sector. Using the results of Sec.~\ref{sec: basis transformation to spin-quadrupole}  and the definition of $g_a$ in Eq.~\eqref{ga}, we have $g_a=g_\mp$ for the spin and quadrupole sectors, respectively. Hence, for the term quadratic in the mean-field currents,
\begin{equation}
    (g_b \pm g_c)f^{bac}j_L^bj_R^c,
\end{equation}
we have $g_b=g_c$, and the term vanishes due to the antisymmetry of the structure constants $f^{abc}$.
The linearized quadratic perturbation is then given by 
\begin{equation}
(g_b \pm g_c)f^{bac}j_L^b\delta j_R^c + (g_b \pm g_c) f^{bac}j_R^c\delta j_L^b. 
\label{quadratic term}
\end{equation}
The SU(3) structure constants $f^{abc}\neq 0$ iff an odd number of indices belong to the set $\{2,5,7\}$, which corresponds to the spin sector. In the following, we use this property to simplify the quadratic terms for the magnetic field and the single-ion anisotropy. 

We start with the magnetic field, which couples to the spin sector. For $a$ in the spin sector, the quadratic term reduces to
\begin{equation}
f^{bac}\frac{B^b}{(1+\delta_-)}(\delta_- \pm \delta_-)(\delta j_R^c  - \delta j_L^c) .
\end{equation}
For $a$ in the quadrupole sector, the quadratic term reduces to
\begin{equation}
    \frac{f^{bac}}{(1+\delta_-)}\left [ (\delta_- \pm \delta_+)B^b\delta j_R^c+ (\delta_+ \pm \delta_-)B^c\delta j_L^b\right ],
\end{equation}
which can be written in terms of $\delta m$ and $\delta j$ using Eq.~\eqref{magnetization-current definition}.

Next, we consider the single-ion anisotropy, which couples to the quadrupole sector. For $a$ in the spin sector, the quadratic term reduces to
\begin{equation}
f^{bac}\frac{B^b}{(1+\delta_+)}(\delta_+ \pm \delta_+)(\delta j_R^c  - \delta j_L^c) .
\end{equation}
For $a$ in the quadrupole sector, the quadratic term reduces to
\begin{equation}
    \frac{f^{bac}}{(1+\delta_+)}\left [ (\delta_+\pm \delta_-)B^b\delta j_R^c+ (\delta_-\pm \delta_+)B^c \delta j_L^b\right ], 
\end{equation}
which can again be written in terms of $\delta m,\delta j$.
\section{Derivation of the Green's Function}
\label{sec: green function derivation from hydrodynamic equations}
In this section, we outline the derivation of the Green's function. 
We begin by transforming the EOMs from the Gell-Mann basis to the spin-quadrupole basis, in which the dynamical correlations are calculated. Then, we derive the Green's function directly from the linearized EOMs, leading to Eq.~\eqref{green function expression}.

In Appendix~\ref{sec: equations of motion auxiliary data} we showed that the quadratic term reduces to a field term. The EOMs therefore contain two types of terms besides the temporal derivatives: a spatial derivative term and a field term.

We begin by discussing the transformation of the former, which carries a coefficient involving $g_a$ (see Eq~\eqref{eq:eoms}). As discussed previously, $g_a=g_-$ and $g_a=g_+$ for $a$ in the spin and quadrupole sectors, respectively. The transformation to the spin-quadrupole basis therefore does not mix $g_{+}$ and $g_-$, leaving the spatial derivative term invariant. 

Next, we derive the transformation of the field term into the spin-quadrupole basis.
We define
\begin{equation}
    \delta\hat{\phi}_{\alpha\beta} = \left(\psi^\dagger_{R,\alpha}\psi_{R,\beta} \pm \psi^\dagger_{L,\alpha}\psi_{L,\beta}\right )-\langle\psi^\dagger_{R,\alpha}\psi_{R,\beta} \pm \psi^\dagger_{L,\alpha}\psi_{L,\beta}\rangle_0, 
\end{equation}
where the expectation value $\langle\cdots\rangle_0$  is taken with respect to the mean-field Hamiltonian, and 
\begin{equation}
\sum_{\alpha,\beta}T^a_{\alpha\beta} \delta\hat{\phi}_{\alpha\beta} \equiv \delta \hat{\phi}^a 
\end{equation}
is equal to $ \delta m^a$ and $\delta j^a$ for $+$ and $-$ , respectively. 
The contribution of the field term to $\partial_t \delta \hat{\phi}^a$, up to a constant prefactor, is
\begin{equation}
\sum_{b,c,\alpha,\beta} f^{bac}B^bT^c_{\alpha\beta} \delta\hat{\phi}_{\alpha\beta}.
\end{equation}
Using $[T^b,T^a]=i\sum_cf^{bac}T^c$, this simplifies to
\begin{equation}
-i\sum_{b,\alpha,\beta}B^b [T^b,T^a]_{\alpha\beta} \delta\hat{\phi}_{\alpha\beta}=  -i \sum_{\alpha,\beta}[O_B,T^a]_{\alpha\beta} \delta\hat{\phi}_{\alpha\beta}, 
\label{field term}
\end{equation}
where we have introduced the field operator,
\begin{equation}
    O_B = \sum_b B^bT^b.
\end{equation}
We denote by $\{V^c\}$ the spin-quadrupole matrix basis, where $c$ labels the spin and quadrupole operators. 
Transforming the EOMs, Eq~\eqref{eq:eoms}, to the spin-quadrupole basis, the linear field term given in Eq.~\eqref{field term} takes the form
\begin{equation}
-i\sum_{\alpha,\beta}[O_B,V^c]_{\alpha\beta}\delta\hat{\phi}_{\alpha\beta} ,
\end{equation}
which is the fluctuation operator contracted with the matrix $[O_B,V^c]$. For example, taking $O_B=-BS^z,V^c=S^y$ yields
\begin{equation}
    iB\sum_{\alpha,\beta}[S^z,S^y]_{\alpha\beta}\delta\hat{\phi}_{\alpha\beta} =B\sum_{\alpha,\beta}S^x_{\alpha\beta}\delta\hat{\phi}_{\alpha\beta},
\end{equation}
which is the fluctuation corresponding to $S^x$, multiplied by the magnetic field strength $B$.

We now proceed to derive Eq.~\eqref{green function expression}. Following Ref.~\cite{wang2022hydrodynamics},
\begin{align}
\partial_t \mathcal{G}^{ab}(x,t;0,0) =& -i\delta(t) \langle[\delta\psi^a(x,t), \delta\psi^b(0,t)]\rangle \nonumber \\ &\quad-i\theta(t) \langle[\partial_t\delta\psi^a(x,t), \delta\psi^b(0,0)]\rangle .
\end{align}
We perform a Fourier transform in space and an inverse Fourier transform in time.
The Fourier transform of the first term on the right-hand side equals $-i\mathcal{F}(k)$, where $\mathcal{F}(k)$ is defined in Eq.~\eqref{F matrix definition}. This term can be calculated using the Kac-Moody commutation relations, Eq.~\eqref{eq: Kac-Moody}.

The second term on the right-hand side is given by
\begin{align}
&\int_{t=0}^\infty dt\,\int_{x=-\infty}^{\infty}dx\,e^{i\omega t-ikx} \langle[\partial_t \delta \psi ^a(x,t), \delta\psi^b(0,0)]\rangle \nonumber \\
&=\int_{t=0}^\infty\int_{x=-\infty}^{\infty} dx\,dt\,e^{i\omega t-ikx} \langle[\mathcal{A}_{ac} \delta \psi ^c(x,t), \delta\psi^b(0,0)]\rangle ,
\end{align}
where the operator matrix $\mathcal{A}$, introduced in Eq.~\eqref{linear eq system form},  encodes the information about the linearized EOMs. 

Since $\mathcal{A}_{ac}$ contains only constants and first derivatives, the Fourier transform is straightforward, and is given by
\begin{equation}
i\mathcal{A}(k)\cdot \mathcal{G}  ,
\end{equation}
where we used the definition of the Green's function,
\begin{equation}
\mathcal{G}_{ab} \equiv (-i)\int_{x=-\infty}^{\infty} dx\,e^{-ikx}\int_{t=0}^\infty dt\,e^{i\omega t} \langle[\delta \psi ^a(x,t), \delta\psi^b(0,0)]\rangle .
\end{equation}
We finally reach
\begin{equation}
-i\omega \mathcal{G}(k,\omega) = -i\mathcal{F}(k) -i(i\mathcal{A}(k)\mathcal{G}(k,\omega)) ,
\end{equation}
which leads to Eq.~\eqref{green function expression} in the main text.

Using the above results, we can calculate the Green's functions for operators in the spin-quadrupole basis. For example,
\begin{equation}
    [S^+, S^-]=[S^x, S^x] + i[S^y, S^x] -  i[S^x, S^y] + [S^y, S^y], 
\end{equation}
and thus the transverse spin Green's function which determines $S^{+-}(k,\omega)$, is given by
\begin{equation}
   \mathcal{G}^\pm_S = \mathcal{G}^{xx}+\mathcal{G}^{yy}+i(\mathcal{G}^{yx}-\mathcal{G}^{xy}) 
\end{equation}
where $\mathcal{G}^{ij}$ for $i,j\in\{x,y\}$ denotes the Green's function associated with the commutator $[\delta m^i(x,t),\delta m^{j}(0,0)]$.

Similarly, the transverse quadrupole Green's function, which determines $Q^{+-}(k,\omega)$, is given by
\begin{equation}
    \mathcal{G}^{\pm}_Q =  \mathcal{G}^{11} +  \mathcal{G}^{22}  + i ( \mathcal{G}^{21}- \mathcal{G}^{12}), 
\end{equation}
where $\mathcal{G}^{ij}$ for $i,j\in\{1,2\}$ denotes the Green's function associated with the commutator $[\delta m^{Q_{i}}(x,t),\delta m^{Q_{j}}(0,0)]$, with $Q_{i,j}$ defined in Eq.~\eqref{eq:quadrupole operators definition}.

\FloatBarrier

\section{Additional Numerical Results in Presence of a Magnetic Field}
\label{sec: magnetic field additional numerical results}

\begin{figure}
    \centering
    \includegraphics[width=\columnwidth]{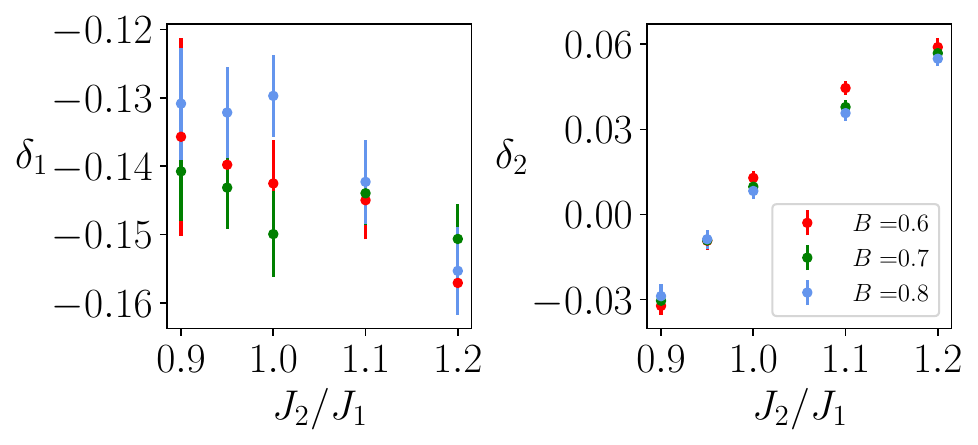}
    \caption{Dimensionless couplings, $\delta_1$ and $\delta_2$, as function of $J_2$, extracted independently for each value of $B$ presented in the main text (Fig.~\ref{fig:spin gap (a) and quadrupole lower-branch shift (b) at k=0 vs. magnetic field}).}
    \label{fig:couplings for all magnetic fields}
\end{figure}

In this Appendix we present a slightly different analysis of the numerical results for the magnetic field case, in which the normalized couplings $\delta_{1,2}$ are extracted independently for each value of $B$ and $J_2$ considered in Sec.~\ref{sec:magnetic field numerical results}.
As discussed therein, the upper upward spin branch and the downward quadrupolar branch, for each $B$ and $J_2$, are first fitted separately to the dispersions in Eqs.~\eqref{eq:spin resonance} and~\eqref{eq:quad resonance}. 
Denoting by $\delta_-^S$ the value of $\delta_-$ obtained from the fit of the spin branch, and by $\delta_2^Q$ the value of $\delta_2$ obtained from the fit of the quadrupole branch, we estimate the couplings to be
\begin{equation}
    \delta_1=\delta_-^S+\delta_2^Q,\quad \delta_2=\delta_2^Q, 
\end{equation}
This choice is motivated by the fact that the spin gap is controlled solely by $\delta_-^S$, while the quadrupole energy shift is determined by $\delta_2^Q$ to leading order in the normalized couplings (see Eqs. ~\eqref{eq: spin spectral gap at $k=0$} and~\eqref{eq: quad shift}).
The resulting couplings are shown in Fig.~\ref{fig:couplings for all magnetic fields}. As can be seen, the couplings obtained at different values of $B$ are consistent with each other. In addition, we see that the value of $\delta_1$ is largely insensitive to $J_2$ in the studied parameter range.

\section{Excitation Branches in Presence of Single-Ion Anisotropy}
\label{sec: single-ion additional analytical results}

In this Appendix, we provide the expressions for the two excitation branches probed by the transverse spin dynamical response in presence of single-ion anisotropy.
These can be written as
\begin{equation}
    (\omega^S_\pm)^2(k) = \frac{D^2}{(\delta_++1)^2}[\tilde{\omega}_{\rm sym}^{2}(k)\pm \tilde{\omega}_{\rm asym}^{2}(k)].
    \label{eq: single-ion branches}
\end{equation}

Denoting $\alpha_{\pm}={\delta_-}\pm1$ and $\beta_{\pm}={\delta_+}\pm1$,  the dimensionless dispersions $\tilde{\omega}_{\rm sym}^{2}$, and $\tilde{\omega}_{\rm asym}^{2}$,
are given by
\begin{align}  
& \tilde{\omega}_{\rm sym}^2(k) =  (1+\delta_-\delta_+) + \nonumber \\
& + \frac{(vk)^2}{2D^2}\big(2+4\delta_++\delta_+^2-2\delta_+^3-\delta_+^4-\delta_-^2- 
   2\delta_-^2\delta_+-\delta_-^2\delta_+^2 \big),
   \label{eq: eps_sym single-ion full}
\end{align}
and
\begin{align}
&\tilde{\omega}_{\rm asym}^{2}(k) = \sqrt{ \tilde{\omega}_1(k) + \tilde{\omega}_2(k)},
\label{eq: eps_asym single-ion full}
\end{align}
where
\begin{align}
\tilde{\omega}_1 = &\frac{1}{4} \alpha_{-}^{2} \beta_{-}^{2} - \frac{1}{2} \alpha_{-} \alpha_{+}\beta_{-} \beta_{+} - \nonumber \\
&-\frac{(vk)^2}{D^2} \alpha_{-}\beta_{+}^{3} \left(\frac{1}{2} \alpha_{+}^{2} - 2 \alpha_{+} \beta_{-} + \frac{1}{2} \beta_{-}^{2}\right) + \nonumber \\
&+ \alpha_{+}\beta_{+}^{2} \left(\frac{1}{4} \alpha_{+} - \frac{1}{2} \frac{(vk)^2}{D^2} \alpha_{-}^{2} \beta_{-}\right),  \nonumber \\
\tilde{\omega}_2 = &- \frac{(vk)^2}{D^2} \beta_{+}^{4} \bigg(\frac{1}{2} \alpha_{+}\beta_{-}- \frac{(vk)^2}{4D^2} \alpha_{-}^{2} \alpha_{+}^{2} +  \nonumber \\
&+ \frac{(vk)^2}{2D^2} \alpha_{-}\alpha_{+}\beta_{-}\beta_{+} - \frac{(vk)^2}{4D^2} \beta_{-}^{2} \beta_{+}^{2}\bigg).
\end{align}

\section{Dynamical Response in Presence of Single-Ion Anisotropy and Magnetic Field}
\label{sec:combining fields}

Since many materials exhibit intrinsic single-ion anisotropy, it is interesting to study the combined effect of this perturbation along with an external magnetic field, which can be tuned experimentally. 
Here, we consider a scenario in which a fairly strong magnetic field is applied in parallel to the axis of a weak single-ion anisotropy, i.e. $\vec{D}\parallel \vec{B}$ and $D \ll B$. 
We show that in this case, the analytical dispersion calculated in absence of single-ion anisotropy, Eq.~\eqref{eq:quad resonance}, captures the observed branches to a good approximation. In particular, the normalized pairing coupling $\delta_2$ can still be extracted from the quadrupolar response using fits to Eq.~\eqref{eq:quad resonance} with good accuracy.

\begin{figure}[ht]
    \centering
    \includegraphics[width=\columnwidth]{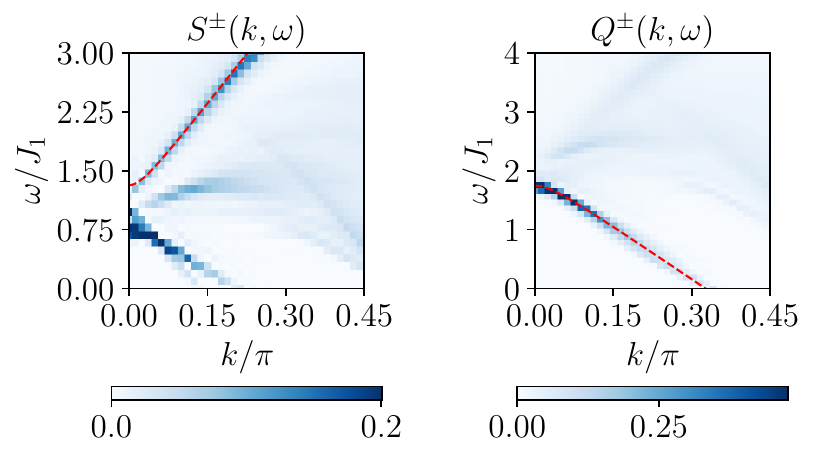}
    \caption{Dynamical spin, $S^{+-}(k,\omega)$, and quadrupolar, $Q^{+-}(k,\omega)$, correlations obtained numerically for the microscopic spin-1 BLBQ model for $B=0.8J_1,J_2=1.1J_1,D=0.1J_1$.}
    \label{fig: Structure Factors for B=0.8,J_2=1.1,D=0.1}
\end{figure}

We consider a regime of $J_2$ and $B$ for which the system remains in the critical phase also in presence of small but finite $D$.
In Fig.~\ref{fig: Structure Factors for B=0.8,J_2=1.1,D=0.1} we show representative spin and quadrupolar dynamical structure factors obtained numerically for $J_2/J_1=1.1$, $B/J_1=0.8$ and $D/J_1=0.1$, with the branches fitted using Eqs.~\eqref{eq:spin resonance} and~\eqref{eq:quad resonance} (ignoring corrections to the dispersion caused by $D$). We obtain $\delta_2 = 0.037(2)$, identical within error bars to the value obtained in the presence of magnetic fields for the same value of $J_2$ and with vanishing single-ion anisotropy, $\delta_2(J_2=1.1J_1,D=0) = 0.040(2)$ (see Fig.~\ref{delta_2 vs. J2 for a magnetic field}). 

\twocolumngrid
\bibliography{references}
\end{document}